\documentclass[]{aa}
\usepackage{graphicx}
\usepackage{txfonts}
\usepackage{natbib}

\bibpunct{(}{)}{;}{a}{}{,}

\newcommand\bb[1] {   \mbox{\boldmath{$#1$}}  }
\newcommand\del{\bb{\nabla}}
\newcommand\bcdot{\bb{\cdot}}
\newcommand\btimes{\bb{\times}}

\begin{document}

\title{MHD simulations of the
  magnetorotational instability in  a shearing box with zero net flux. I. The issue of convergence}
\author{ S\'ebastien Fromang and John Papaloizou }

\offprints{S.Fromang}

\institute{Department of Applied Mathematics
and Theoretical Physics, University of Cambridge, Centre for
Mathematical Sciences, Wilberforce Road, Cambridge, CB3 0WA, UK  \\ \email{S.Fromang@damtp.cam.ac.uk}}

\date{Accepted; Received; in original form;}

\label{firstpage}

\abstract
{}
{We study the properties of MHD turbulence driven by the
  magnetorotational instability (MRI) in accretion
  disks.  To do this we  perform a series of numerical simulations  for which
  the resolution is gradually increased.}
{We adopt the local shearing box  model and focus on the special case
  for which
  the initial magnetic flux threading the disk vanishes. We employ the finite
  difference code ZEUS to evolve the ideal MHD equations.}
{Performing a set of numerical simulations in a fixed computational
domain  with increasing resolution, we
  demonstrate that turbulent activity decreases as resolution 
  increases. The highest resolution considered is $256$ grid cells per
  scale height. We quantify
  the turbulent activity by measuring the rate of angular momentum
  transport  through evaluating  the standard $\alpha$ parameter. We find
  $\alpha=0.004$ when $(N_x,N_y,N_z)=(64,100,64)$,
  $\alpha=0.002$ when $(N_x,N_y,N_z)=(128,200,128)$ and
  $\alpha=0.001$ when $(N_x,N_y,N_z)=(256,400,256)$. This steady
  decline is
  an indication that numerical dissipation, occurring at
  the grid scale  is an important determinant of the saturated  form
  of the MHD 
  turbulence. Analysing the results in Fourier space, we demonstrate
  that this is due to the MRI forcing significant flow  energy all the
  way down to the
  grid dissipation scale. We also use our results to study the
  properties of the numerical dissipation in ZEUS. Its amplitude is
  characterised by the magnitude of an effective magnetic Reynolds number
  $Re_{M}$ which increases from $10^4$ to $10^5$  as the
  number of grid points is increased from $64$ to $256$ per scale height.}
{The simulations we have carried out do not produce results that are independent
of the numerical dissipation scale, even at the highest resolution studied.
Thus it is important to use physical dissipation, both viscous and
resistive, and to  quantify contributions from numerical effects, when
performing numerical simulations of MHD turbulence with zero net
  flux in accretion disks at the resolutions normally considered.}

\authorrunning{S.Fromang \&  J.Papaloizou}
\titlerunning{MHD turbulence in accretion disks. I. Convergence}
\maketitle

\section{Introduction} 
\label{intro}

A long standing issue in accretion disk theory  has been to identify the
source of anomalous transport of angular momentum. To date, the most
likely mechanism is believed to be the magnetorotational instability
\citep[MRI;][]{balbus&hawley98} which simply requires a weak magnetic
field and a radially decreasing angular velocity to operate in a highly conducting disk. 
Appropriate conditions are readily realised in many 
 astrophysical accretion disks and the linear instability
grows on dynamical timescales. Its nonlinear evolution has been widely
studied since the early 1990's.  Local simulations using the shearing box model
\citep{hawleyetal95,brandenburgetal95}   were found to give rise to MHD
turbulence  with an associated  rate of angular momentum transport
compatible with the observations, with $\alpha$, the standard
parameter in standard disk theory \citep{shakura&sunyaev73}, being in the range
$10^{-3}$--$0.1$ depending on the geometry of the magnetic field.

In this paper we focus on MHD simulations in a shearing box threaded
 by zero net magnetic flux initially and use the operator split
 code ZEUS \citep{hawley&stone95}. The  first detailed consideration
 of this case was by \citet{hawleyetal96}. The potential importance
 for angular momentum transport in accretion disks is that it offers
 the possibility of   local turbulence coupled with genuine   dynamo
 action.  If such activity can be maintained, it would be independent
 of any imposed magnetic field and   being local,  independent of
 distant boundary conditions.  Accordingly this would be a robust
 outcome of the MRI, providing a guaranteed level of
 transport. However, some recent results \citep{gardiner&stone05b}
 indicate that the  saturated  turbulent state is sensitive to
 numerical resolution. Accordingly, issues remain as to whether a
 numerically converged saturated turbulent state can be achieved.  It
 is important to note that should such simulations ultimately yield
 negligible or zero transport, the MRI can still operate to produce
 sustained turbulence and transport, but more attention would have to
 be paid to imposed fields and boundary conditions in the context of
 global simulations which have been  and are being currently  carried
 out \citep{hawley01,steinacker&pap02,fromang&nelson06}.

The plan of the paper is as follows: In section \ref{run_setup} we
describe the computational set up for the simulations we performed
and algorithm used. We then go on to present results for  typical
runs.  In section~\ref{resol_effect_sec} we discuss the effect of
resolution on the results and in section~\ref{Fspc} we discuss the
power spectra associated with the saturated turbulent states and use
these to show that significant flow energy is always driven to the
smallest numerically realisable scales. Thus results remain dependent
on resolution at the highest resolution studied.  We then go on to
discuss these results and their implications for understanding the non
linear outcome of the MRI  in  section \ref{discussion_section} and
give our conclusions in section \ref{conclusion_section}.

\section{Initial conditions and run setup}
\label{run_setup}

In this paper  we solve the ideal MHD equations in a shearing box
  \citep{goldreich&lyndenbell65} using ZEUS \citep{hawley&stone95}.
  To do this we adopt a  Cartesian coordinate system $(x, y, z)$  with
  unit vectors $(\bb{i},\bb{j},\bb{k})$ pointing in the directions of
  the coordinate axes. The vertical direction is defined by $\bb{k}$
  and the azimuthal direction by $\bb{j}.$ The frame  rotates with the
  angular velocity  of a free  particle in circular orbit  at the
  centre of the box and the origin of the coordinate system
  $\bb{\Omega}=\Omega \bb{k}$.  In this frame the basic equations can
  be written as

\begin{eqnarray}
\frac{\partial \rho}{\partial t} + \del \bcdot (\rho \bb{v})  &=&  0 \,\label{contg} , \\
\rho \frac{\partial \bb{v}}{\partial t} + \rho ( \bb{v} \bcdot \del )
\bb{v}   + 2 \rho
\bb{\Omega} \times \bb{v}   &=&  - \del P + \frac{1}{4\pi} (\del \btimes \bb{B})
\btimes \bb{B} , \\
\frac{\partial \bb{B}}{\partial t}  &=&  \del \btimes ( \bb{v} \btimes
\bb{B}  ) \, \label{induct} .
\label{shearing_sheet_eq}
\end{eqnarray}
 Here $\rho$ stands for the gas
density, $\bb{v}$ for the velocity, $\bb{B}$ for the magnetic field
 and $P$ for the pressure.

To close the system written above, one need to specify the pressure
through an equation of state. For reasons of simplicity,   throughout
this paper we adopt an isothermal equation of state for which
\begin{equation}
P=\rho c_0^2 \, .
\end{equation}
As usual, the ratio between the speed of sound $c_0$ and the angular
frequency $\Omega$ can be used to  define a disk scale height $H$.

Given the above framework, a simulation is defined once the size of
the box, the resolution and the initial geometry and strength of the
magnetic field are chosen. In this paper,  for the most part, we
consider computational boxes of size $(L_x,L_y,L_z)\equiv (H,\pi
H,H)$, although we  have also considered  boxes of size $(H,2 \pi
H,H)$ in order to  compare our  simulations with  those published in
the literature previously. The resolutions  used vary from
$(N_x,N_y,N_z)=(64,100,64)$ to $(N_x,N_y,N_z)=(256,400,256)$. We will
measure times in units of the orbital period, $T=2\pi /\Omega$. As
mentioned in the introduction, we focus exclusively on the special
case in which no net magnetic flux threads the box (in the vertical or
azimuthal directions) initially. The magnetic field at the start of
the simulation is purely vertical and defined as follows:
\begin{equation}
B_z=B_0 \sin (2\pi x/H)
\end{equation}
where $B_0$ is set such that the volume average ratio between thermal
and magnetic pressure $<\beta>$ equals $400$. We checked, however,
that the saturated state of the turbulence depends neither on that
value nor on the geometry of the field provided the net flux remains
zero. At the beginning of each simulations, random velocity
fluctuations of small amplitude are applied to  an initial state with
uniform  gas density and velocity that is entirely due to the
background Keplerian shear and takes the form $\bb{v} = (0, -3\Omega
x/2, 0).$ All the simulations performed here had the Courant number $C
= 1/2.$

Following standard practise \citep{hawleyetal95} strictly periodic
boundary conditions are applied in $y$ and $z$  while boundary
conditions that are periodic in shearing coordinates are applied in
$x$. The latter require some care as they might introduce
  spurious numerical artefacts. Here, we
  applied these shearing box boundary conditions directly to the
  magnetic field. Although this procedure safely conserves the mean
  radial magnetic field threading the box to within round--off error,
  the mean y and z components of the field are only conserved to
  within truncation error. Thus there is a possibility of having long
  term accumulation of azimuthal or vertical magnetic fluxes,
  which would increase the MRI--induced turbulent activity in the
  box. In the following, we will therefore monitor the time variation
of the mean y and z components threading the computational domain
during the simulations.

\subsection{Run parameters}
\begin{table*}[t]\begin{center}\begin{tabular}{@{}ccccccc}\hline\hline
Model & Box size & Resolution & Run time (in orbits) & $\alpha_{Rey}$
& $\alpha_{Max}$ & $\alpha$ \\
\hline\hline
FS64 & $(H,2\pi H,H)$ & $(64,200,64)$ & $300$ & $1.8 \times 10^{-3}$ &
$4.2 \times 10^{-3}$ & $5.9 \times 10^{-3}$ \\
\hline
STD64 & $(H,\pi H,H)$ & $(64,100,64)$ & $1000$ & $9.4 \times 10^{-4}$
& $3.2 \times 10^{-3}$ & $4.1 \times 10^{-3}$ \\
STD128 & $(H,\pi H,H)$ & $(128,200,128)$ & $250$ & $5.0 \times 10^{-4}$
& $1.7 \times 10^{-3}$ & $2.2 \times 10^{-3}$ \\
STD256 & $(H,\pi H,H)$ & $(256,400,256)$ & $105$ & $2.4 \times 10^{-4}$ & $8.1 \times 10^{-4}$ & $1.1 \times 10^{-3}$ \\
\hline
STD64a & $(H/2,\pi H/2,H/2)$ & $(64,100,64)$ & $120$ & $3.3 \times 10^{-4}$ & $1.4 \times 10^{-3}$ & $1.7 \times 10^{-3}$ \\
\hline
LB64 & $(2H,2\pi H,H)$ & $(128,200,64)$ & $120$ & $1.5 \times 10^{-3}$ & $3.6 \times 10^{-3}$ & $5.2 \times 10^{-3}$ \\
LB128 & $(2H,2\pi H,H)$ & $(256,400,64)$ & $105$ & $8.3 \times 10^{-4}$ & $2.1 \times 10^{-3}$ & $2.8 \times 10^{-3}$ \\
\hline\hline
\end{tabular}
\caption{Properties of the runs described in this paper: The first column
  gives  the model label, while the next three  columns
  give the size of the
  computational domain $(L_x,L_y,L_z),$ the resolution $(N_x,N_y,N_z)$
  and the time (in orbits) for
  which the simulation was run respectively. The fifth to seventh columns indicate the
  rate of angular momentum transport by giving the volume and time
  averaged value of the Reynolds, Maxwell and total stresses, normalised
  by the initial thermal pressure, respectively (note that these
  values are obtained in each
  case by averaging the results from $t=40$ orbits until the end of
  the simulation).}
\label{zero_diss_prop}
\end{center}
\end{table*}

The details of the runs we performed are given in Table
\ref{zero_diss_prop}. The first column gives the simulation  label.
 The second and third column give  the box dimensions
$(L_x,L_y,L_z)$ and the resolution $(N_x,N_y,N_z).$ 
The fourth column gives the simulation duration 
in orbital times. Finally, the last three columns give time
averaged values of the Reynolds, Maxwell and total stresses,
normalised by the initial thermal pressure $P_0$. They measure the rate of angular
momentum transport and as usual 
are respectively defined through
\begin{eqnarray}
\alpha_{Rey}&=&\frac{T_{r\phi}^{Rey}}{P_0}=\frac{1}{P_0}\langle \rho
(v_x-\bar{v_x})(v_y-\bar{v_y})\rangle \, , \\
\alpha_{Max}&=&\frac{T_{r\phi}^{Max}}{P_0}=\frac{1}{P_0}\left \langle -{B_x B_y\over 4\pi}\right \rangle \, , \\
\alpha&=&\alpha_{Rey}+\alpha_{Max}=\frac{T_{r\phi}^{Max}+T_{r\phi}^{Rey}}{P_0} \, .
\end{eqnarray}
In these equations, ${\overline{v}_x}$ and ${\overline{v}_y}$ are averages
over  $y$ and $z$ of the velocity components  in the $x$ and $y$
directions respectively while the angled brackets denote a volume average.
 Note that because of the conservation of mass $P_0 = \langle P\rangle.$

\subsection{Standard runs}

\begin{figure}
\begin{center}
\includegraphics[scale=0.5]{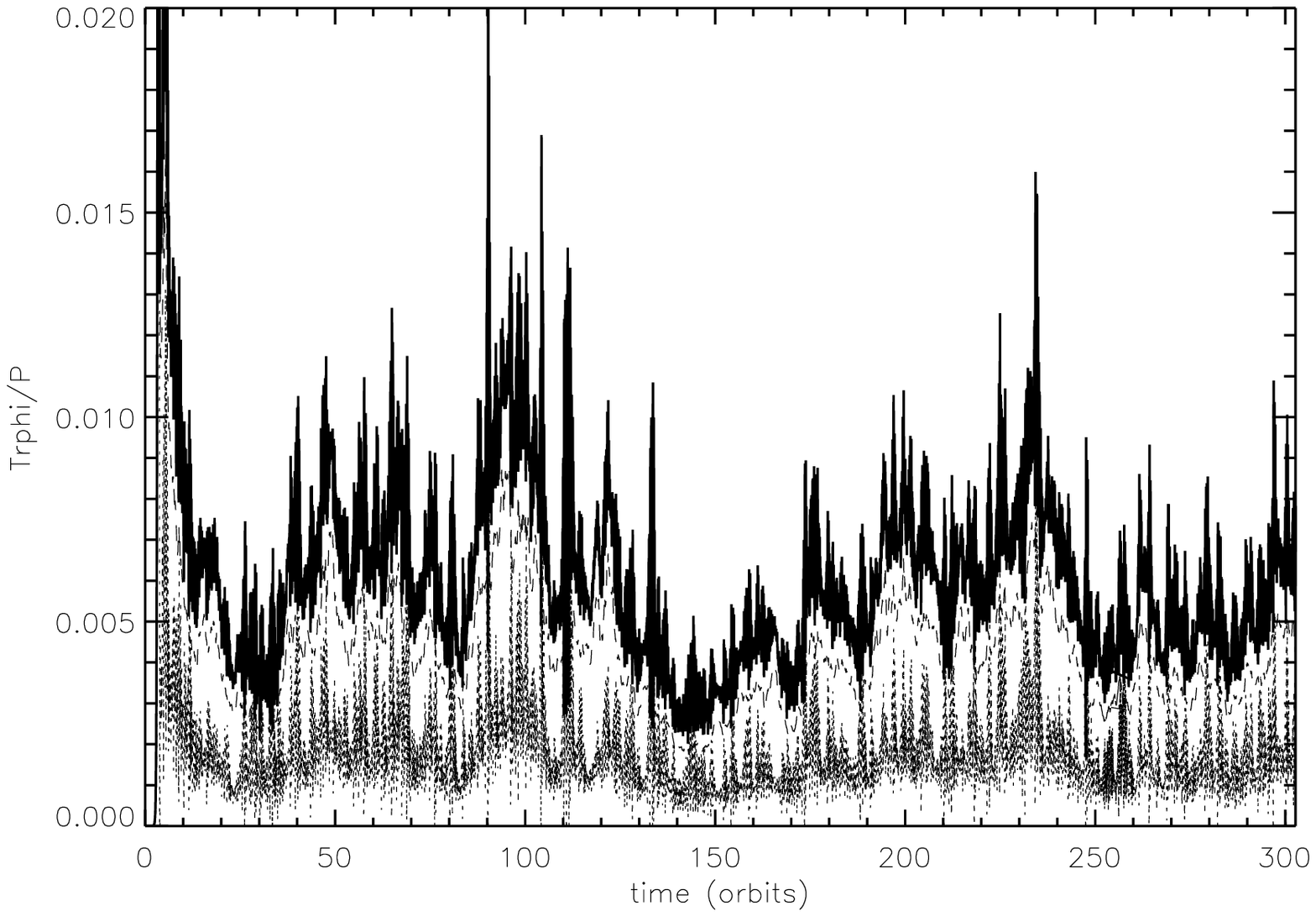}
\includegraphics[scale=0.5]{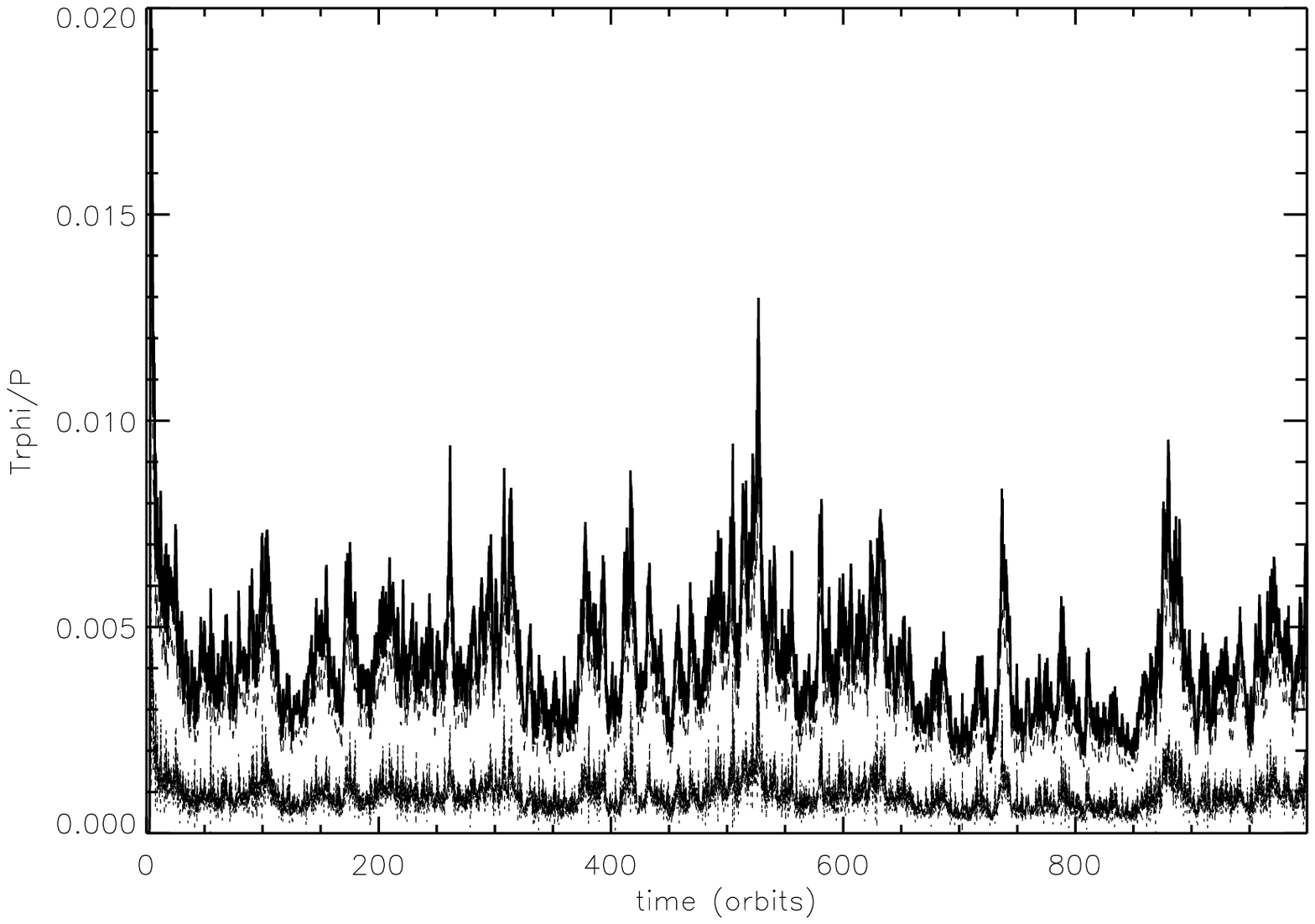}
\caption{Time history of the stress parameters $\alpha_{Rey}$, $\alpha_{Max}$
  and $\alpha$ for the runs FS64 ({\it upper panel}) and STD64 ({\it
  lower panel}). For each plot, the dotted
  curve corresponds to the Reynolds stress, the dashed curve
  corresponds to the Maxwell stress while the solid curve is the
  sum of the two. All of these are normalised by the  initial thermal pressure.}
\label{mri_time_hist_64}
\end{center}
\end{figure}

To make a connection with previously published results, we first perform
two runs with a moderate resolution. Both have $64$ grid cells in
the $x$ and $z$ direction. The first, labelled FS64, uses a box size
$(L_x,L_y,L_z)=(H,2\pi H,H)$ and $200$ cells in the y direction. It
is almost identical to one of the runs presented by
\citet{flemingetal00}. The only difference is that these authors
use an adiabatic equation of state
and a somewhat lower resolution in the $y$ direction. Model FS64 is
compared with model STD64 in which the
size of the computational box is halved in the $y$ direction. To maintain the
same effective resolution, $N_y$ is
also decreased to $100$. Thanks to the improved  computational resources that
have become available in the last few years, models FS64 and STD64 have
 been run for $300$ and $1000$ orbits respectively.

The time history of $\alpha_{Rey}$, $\alpha_{Max}$ and $\alpha$ are
shown for both runs FS64 and STD64 in figure~\ref{mri_time_hist_64}
respectively on the upper and lower panels. In each case, the dotted
line represents $\alpha_{Rey}$, the dashed line shows
$\alpha_{Max}$ while the solid line corresponds to $\alpha$, the
sum of the two. Both runs display the characteristic signatures of the
MRI: an initial growth during the first few orbits due to the linear
instability, a decrease of the stress after reaching a maximum as the
linear instability breaks down into MHD turbulence and finally attainment of a
saturated quasi steady state phase characterised by 
outward angular momentum transport for the remainder of the
simulation. As observed in most simulations of this type, most
of the transport is due to the contribution of the Maxwell
stress. Both runs show sustained MHD turbulence for
hundreds of orbits, in agreement with earlier studies \citep{sanoetal04},
with significant fluctuations occurring on both short (less than $1$ orbit) and long
timescales (more than $10$ orbits). Time--averaged values of the
stresses between $t=40$ and the end of the  simulation, are
given for both runs in Table~\ref{zero_diss_prop}. For model FS64, we
find values almost identical to those reported by
\citet{fleming&stone03} in their Table~2. For model STD64, we find
that the transport is weaker by about $30\%$, as  for this model $\alpha=4.1 \times
10^{-3}$ whereas  $\alpha = 5.9 \times 10^{-3}$ in model FS64. The
difference is due to the smaller box size in model STD64, as this is
the only difference between the two simulations. Although this
relation between the box size and 
turbulent activity is not yet well understood, it was already noted in
earlier calculations of the shearing box \citep{hawleyetal96}.
 
In conclusion, models FS64 and STD64 demonstrate  good
agreement between our results and previously published
calculations. The next step, which is the main goal of this paper, is
to check the convergence of these results when resolution is
increased. Given the large computational cost associated with well
resolved simulations, it is necessary to choose one particular box size for these
runs  despite the differences between the runs FS64 and
STD64 mentioned above. To reduce the
computational burden, we adopt the smaller box
for the remainder of this paper, as  the computing time is reduced  by a
factor of two. The consequences of changing the box size will be
  briefly considered in section~\ref{scaling_section}.

\section{The effect of resolution}
\label{resol_effect_sec}

\begin{figure}
\begin{center}
\includegraphics[scale=0.5]{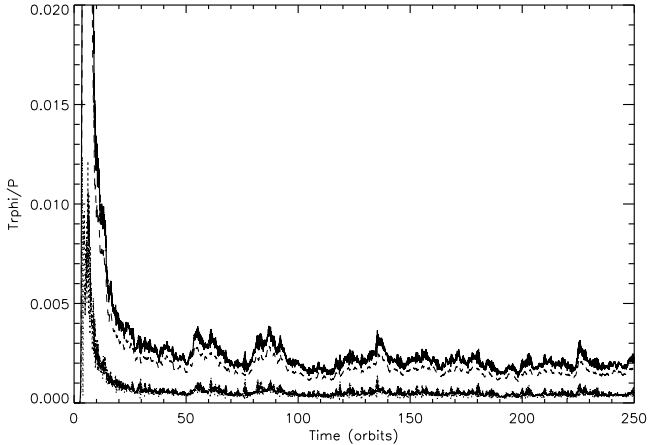}
\includegraphics[scale=0.5]{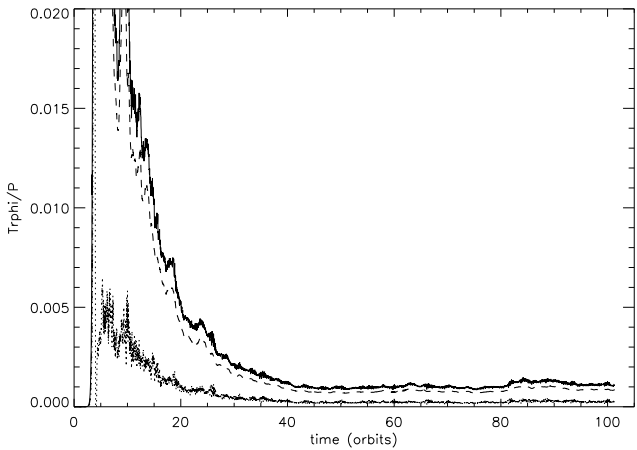}
\caption{Same as figure~\ref{mri_time_hist_64} but for the runs STD128
  ({\it upper panel}) and STD256 ({\it lower panel}). Note the overall
  decreasing turbulent activity as the resolution is increased.}
\label{increase resol}
\end{center}
\end{figure}


\begin{figure}
\begin{center}
\includegraphics[scale=0.5]{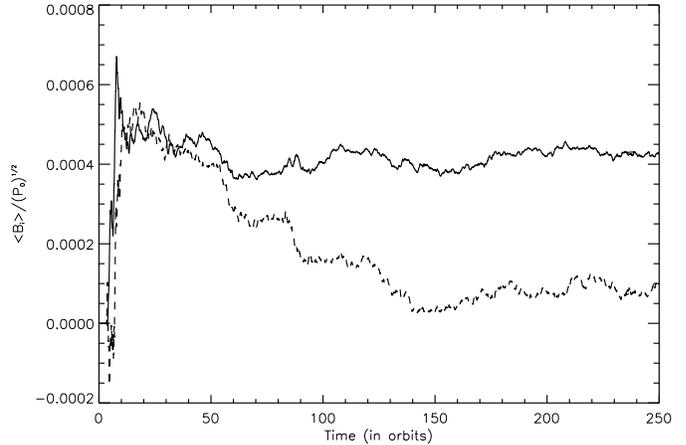}
\caption{Time history of the mean azimuthal ({\it solid line}) and
  vertical ({\it dashed line}) magnetic field threading the box in
  model STD128, normalised by $P_0^{1/2}$. Both remain small enough no
  to affect the long term evolution of the simulations.}
\label{fluxes_var}
\end{center}
\end{figure}

In this section, we study the influence of resolution on the results
of model STD64 by doubling (model STD128) and quadrupling (model
STD256) the number of cells in each coordinate direction. All other
parameters are kept identical to those  of  STD64. The time history of
the normalised  stresses  is shown in figure~\ref{increase resol} for
models STD128 and STD256 in the upper and lower panels  (using the
same conventions as in figure~\ref{mri_time_hist_64}) respectively. It
is clear from figure~\ref{increase resol}  that $\alpha_{Max}$  and
$\alpha_{Rey}$   both decrease as resolution is increased. From
Table~\ref{zero_diss_prop}, the time averaged value of $\alpha$ in
model STD128 is $2.2 \times 10^{-3}$ while it is $1.1 \times 10^{-3}$
in model STD256. In other words, in going from $64$ grid points to
$256$ grids points per scale height, the turbulent activity decreases
by approximately a factor of two each time the resolution  increases
by a factor of $2$. Figure~\ref{increase resol}  also shows that the
amplitude of the fluctuations of the stresses tends to decrease
as the  resolution is increased. This could be a signature of the
decreasing importance of channel flows that have been suggested as
being responsible for these fluctuations \citep{sano07}, as resolution
is increased.

Model STD256 is run for $105$ orbits. The lower resolution
  models STD64 and STD128 show that this is enough to
  get a good estimate of the stresses. Indeed, averaging
  $\alpha$ in these two models between $t=40$ and $t=105$, we
  respectively found $\alpha=4.3 \times 10^{-3}$
  and $2.5 \times 10^{-3}$, which are close to the values quoted
  in Table~\ref{zero_diss_prop} and obtained by averaging over much
  longer periods. Averaging the stresses over $105$
  orbits should therefore be enough to get a good estimate of
  $\alpha$.

Next, we turn our attention to possible problems induced by
  the shearing box boundary conditions. To check whether y and z mean
  magnetic field are created in the box, we plot in
  figure~\ref{fluxes_var} the time history of both in model STD128,
  normalised by $P_0^{1/2}$. It shows no systematic increasing accumulation of
  net flux in the computational domain, despite the imperfect nature
  of the shearing box boundary conditions. Furthermore, the absolute
  value of both components is always very small. Their maximum
  strength during the simulation, expressed in terms of effective beta
  values (defined as $\beta_i=8\pi P_0/\!<\!\!B_i\!\!>^2$), is respectively
  $\beta_y=5.4 \times 10^6$ and $\beta_z=4.3 \times 10^6$ for
  the y and z components. For the resolution we are using, this is
  far too small a field strength to have any effect on the saturated
  state of the turbulence (the wavelength of the most unstable MRI
  mode for such weak fields is always smaller than a grid cell). We
  performed the same checks for models STD64 and 
  STD256. For the former, the maximum values of the mean field
  components during
  the simulations corresponded to $\beta_y=2.3 \times 10^6$ and $\beta_z=1.5
  \times 10^7$. For the later, we obtained $\beta_y=3.8 \times 10^8$
  and $\beta_z=1.7 \times 10^8$. All these values indicate that the
  boundary conditions have no effect on the results.

\begin{figure*}
\begin{center}
\includegraphics[scale=0.28]{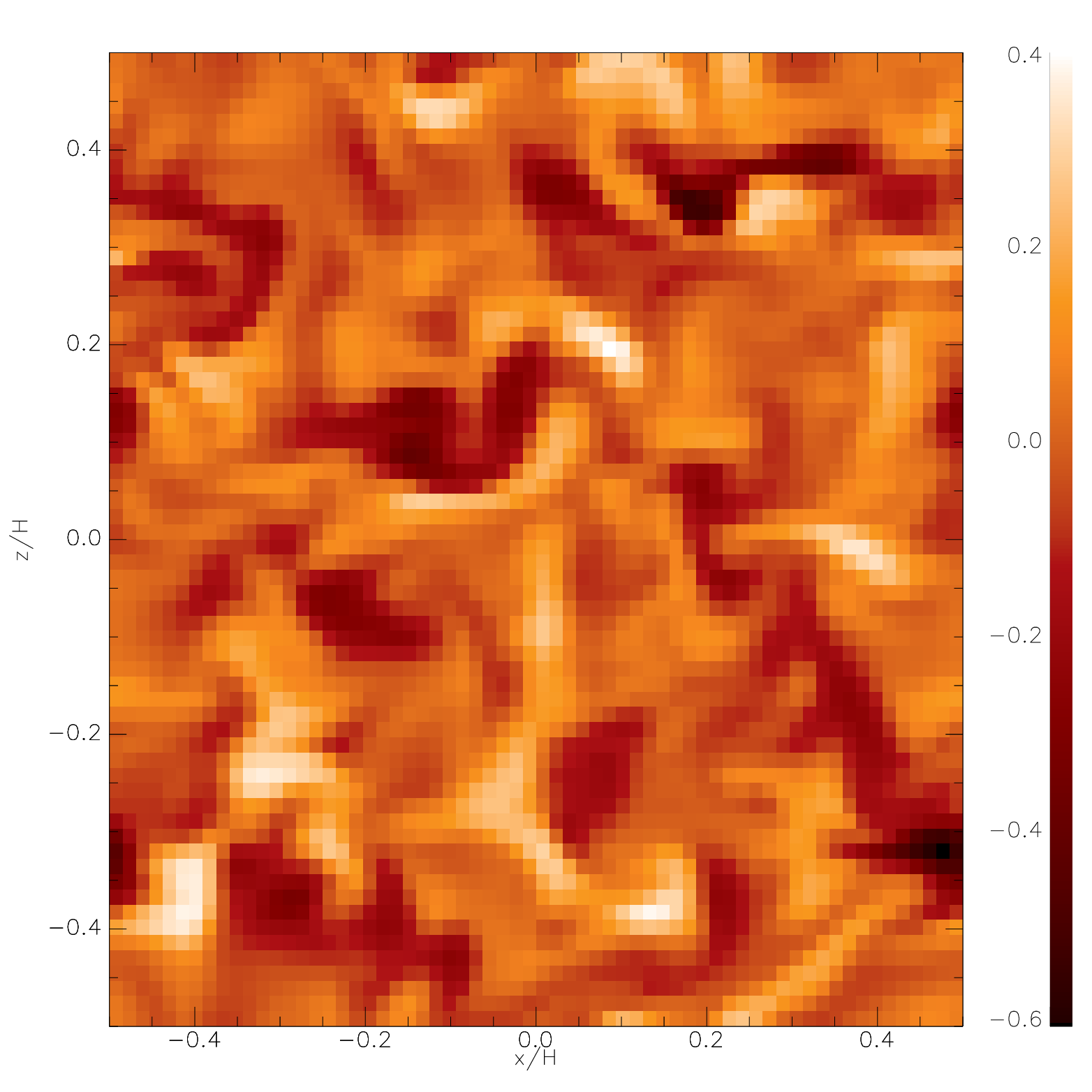}
\includegraphics[scale=0.28]{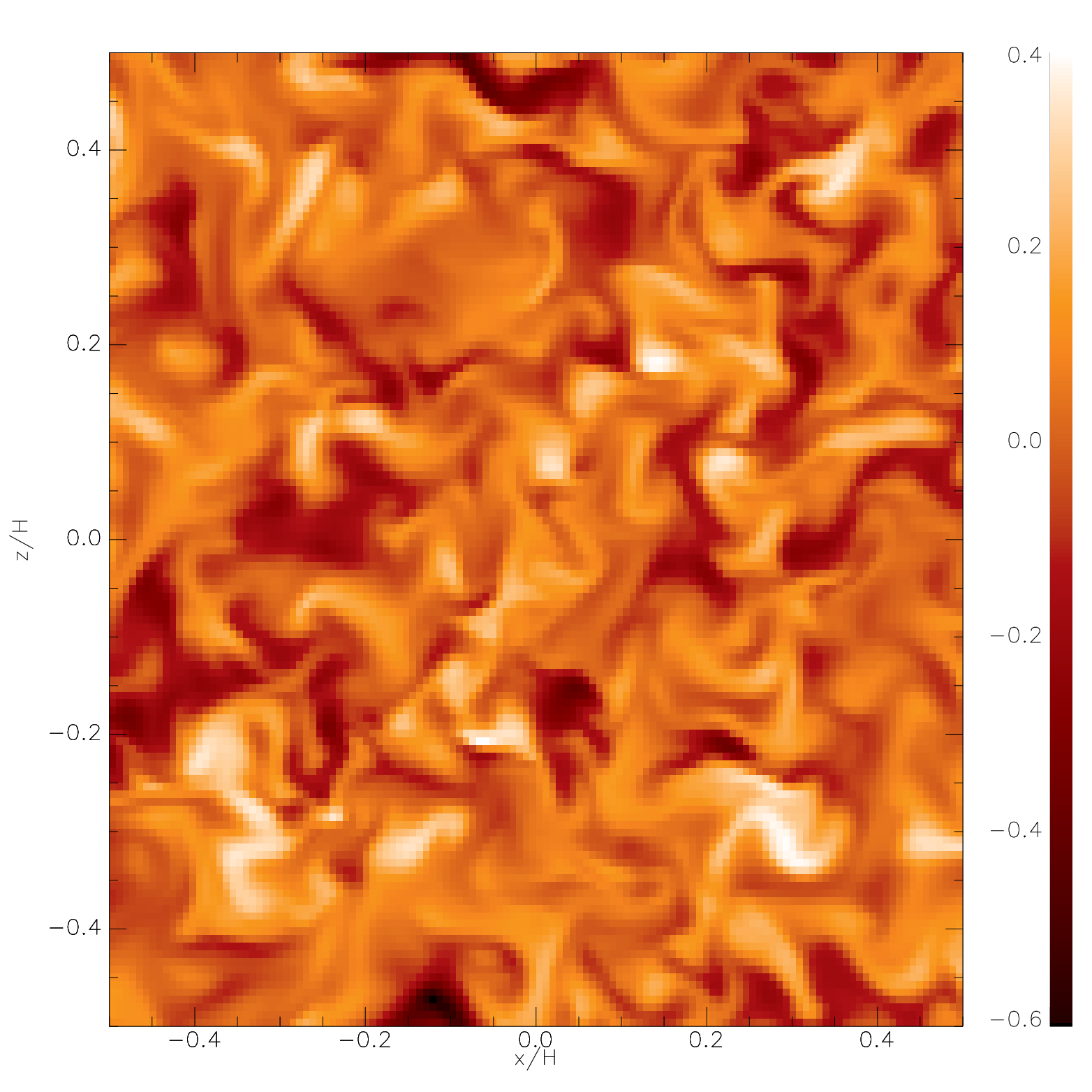}
\includegraphics[scale=0.28]{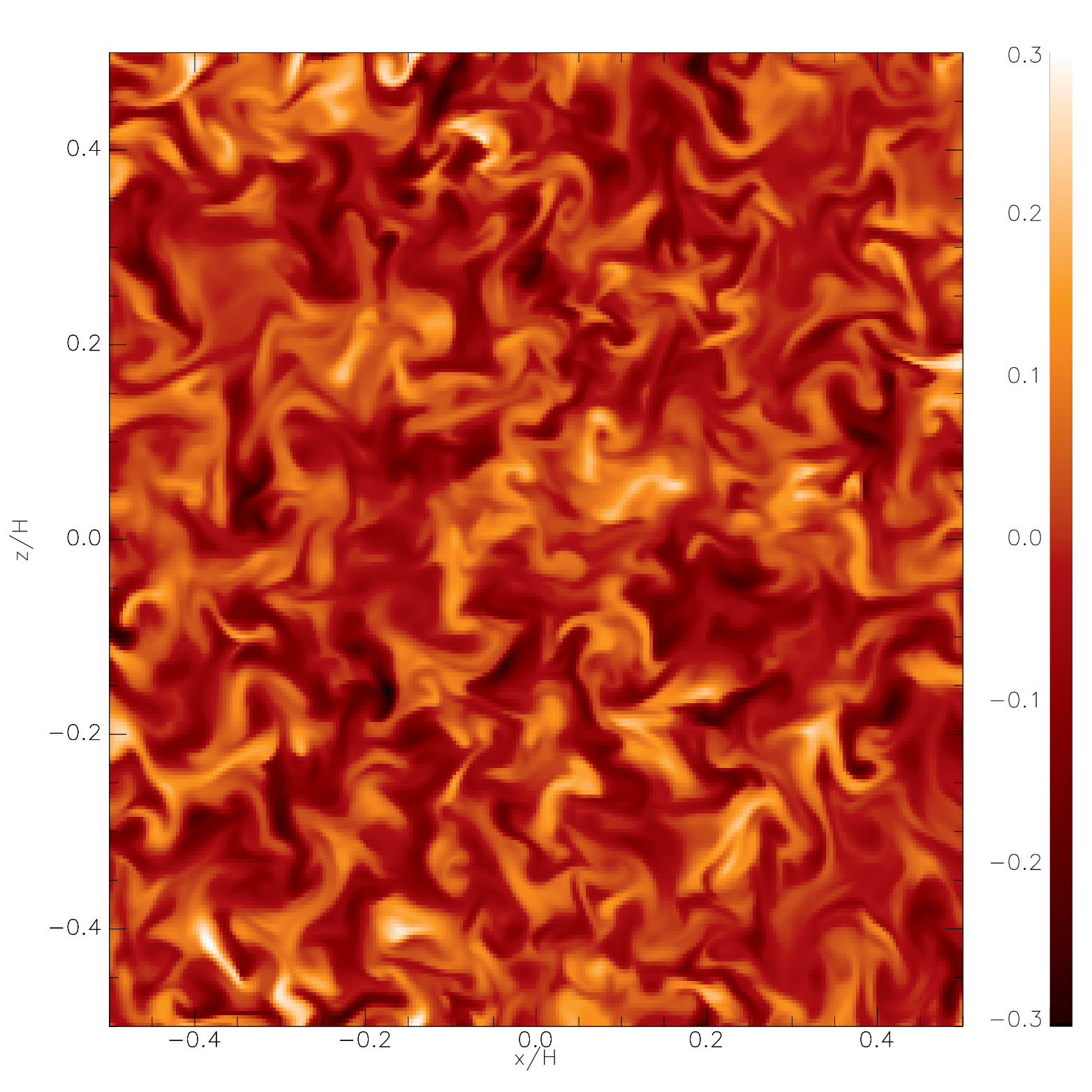}
\caption{Snapshots showing contours of the y--component of the magnetic field in
  the  $(x,z)$ plane $(y=0)$ for the runs STD64 ({\it left panel}), STD128
  ({\it middle panel}) and STD256 ({\it right panel}). Smaller and smaller scale
  features in the magnetic field are seen as the resolution of
  the simulation increases.}
\label{compar_by}
\end{center}
\end{figure*}

\begin{figure}
\begin{center}
\includegraphics[scale=0.5]{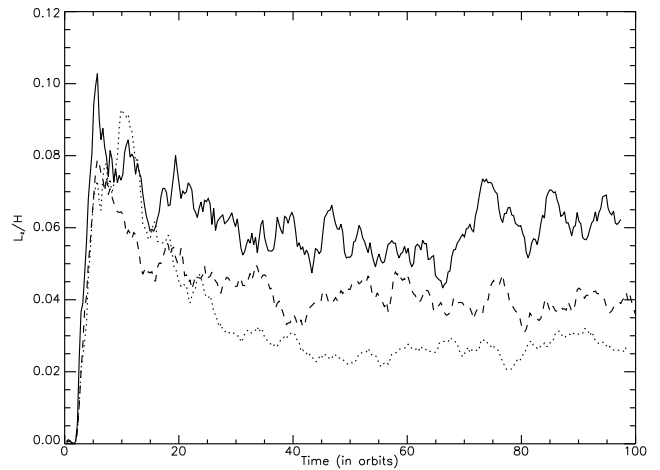}
\caption{Time history of the correlation length in the $z$ direction of $B_y$ 
   for models STD64 ({\it solid line}), STD128 ({\it dashed
  line}) and STD256 ({\it dotted line}). As the resolution is
  increased, $L_z(B_y)$ decreases, as might be expected from
  the appearance of the snapshots
  shown in figure~\ref{compar_by}.}
\label{length_by}
\end{center}
\end{figure}
Being confident that the models we present are not significantly
  affected by
  the boundary conditions, we now turn to a more detailed analysis of their
  properties. To illustrate the changes in the structure of the flow as
resolution is increased, figure~\ref{compar_by} provides snapshots of the
structure of the magnetic field. From left
to right, contours of $B_y$ in the $(x,z)$ plane $(y=0)$  are
given for models STD64
({\it left panel}), STD128 ({\it middle panel}) and STD256 ({\it right
  panel}). As the  resolution is increased, smaller and smaller scale 
structure  becomes apparent. The only limitation on the smallness of the scale
appears to be  due to  finite resolution.
 It is possible to make this statement  more
quantitative by computing a vertical correlation length for $B_y$. Following
\citet{lesur&longaretti07}, we define this correlation length  through
\begin{equation}
L_z(B_y)=\left<\frac{\int  \int B_y(x,y=0,z) B_y(x,y=0,z') dz'dz }{\int
  B_y^2(x,y=0,z) dz }\right> \, .
\label{correl_length}
\end{equation}
In this definition, the symbol angled brackets denote an average over the $x$
direction.  The time
history of $L_z(B_y)$ is  plotted in
figure~\ref{length_by},
using a solid line for model STD64, a dashed line for model STD128 and
a dotted line for model STD256.
 Because the raw data were very noisy,
the initial curves were smoothed using a window of about $3$ orbits,
corresponding to $10$ snapshots. Figure~\ref{length_by} confirms the
contraction of scale apparent from the trends
seen in  figure~\ref{compar_by}: after about $40$
orbits, $L_z(B_y)$ reaches a quasi steady  value which decreases as
resolution is increased. Averaging the results in time from $t=40$
orbits until the end of the run, we obtained $L_z(B_y)/H=0.06$ for model
STD64, $0.04$ for model STD128 and $0.025$ for model STD256. For each
model, these numbers indicate that  typical size scale  for  structures
in $B_y$ is a few grid cells. For model STD64,
$L_z(B_y)$ corresponds to $3.8$ grid cells, while it is equal to $5$ grid
cells for model STD128 and $6.5$ grid cells for model STD256. Similar
values are obtained when calculating a correlation length in the
direction $x$, defined using an equation similar to Eq.~(\ref{correl_length}).

All of these results indicate that the saturated state of  MHD
turbulence in these simulations is governed by the numerical
dissipation of the code. It is therefore important to understand in a
more detailed way the dissipation in ZEUS (and by extension
any MHD code without specified diffusivities that relies
on numerical dissipation to bound the size scale from below) and why
and how it affects the
results. In the following section, we use Fourier analysis to address this issue.

\section{Fourier analysis}\label{Fspc}

\subsection{Power spectra}
\label{power_spec_sec}

\begin{figure}
\begin{center}
\includegraphics[scale=0.45]{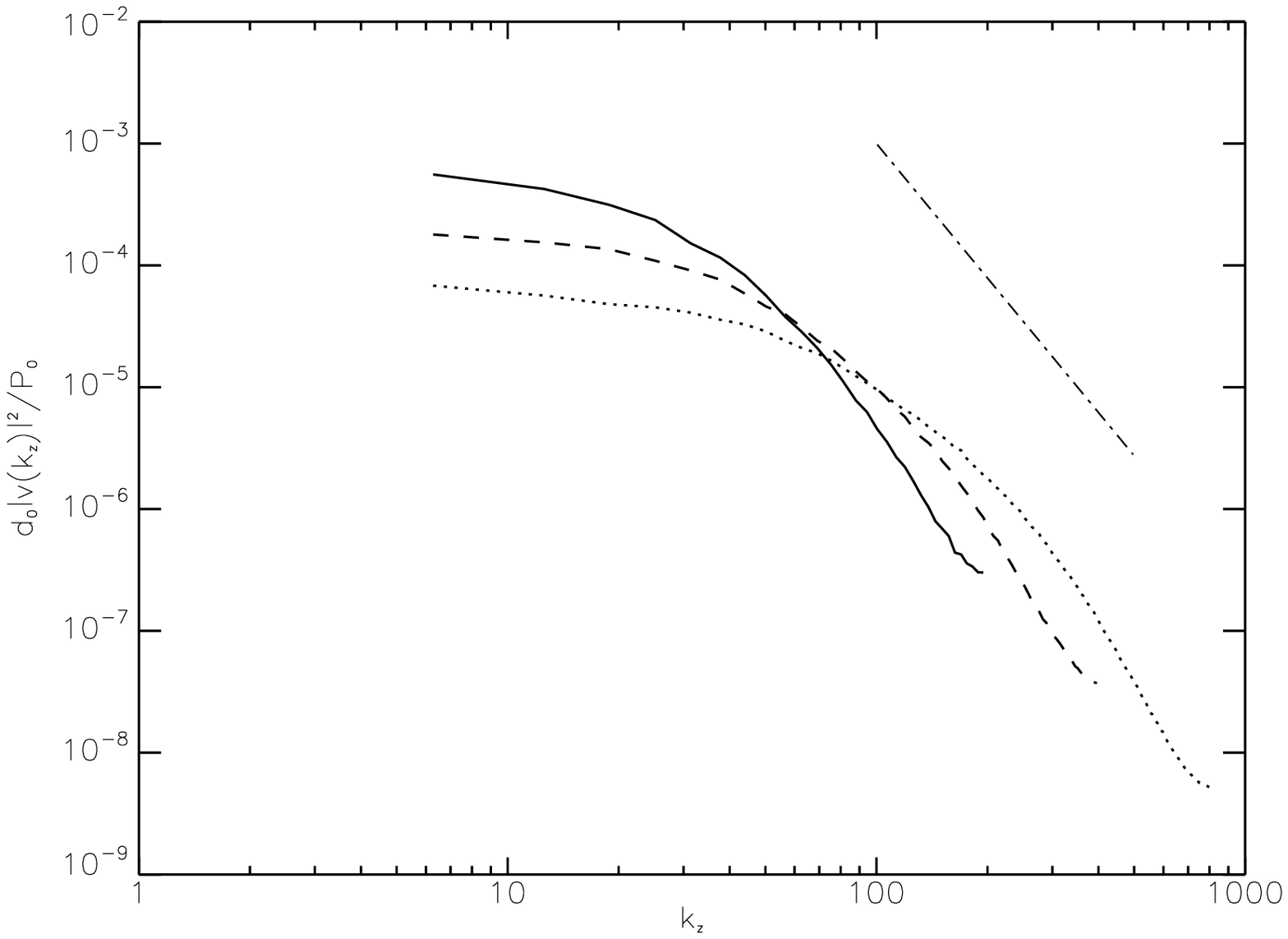}
\includegraphics[scale=0.45]{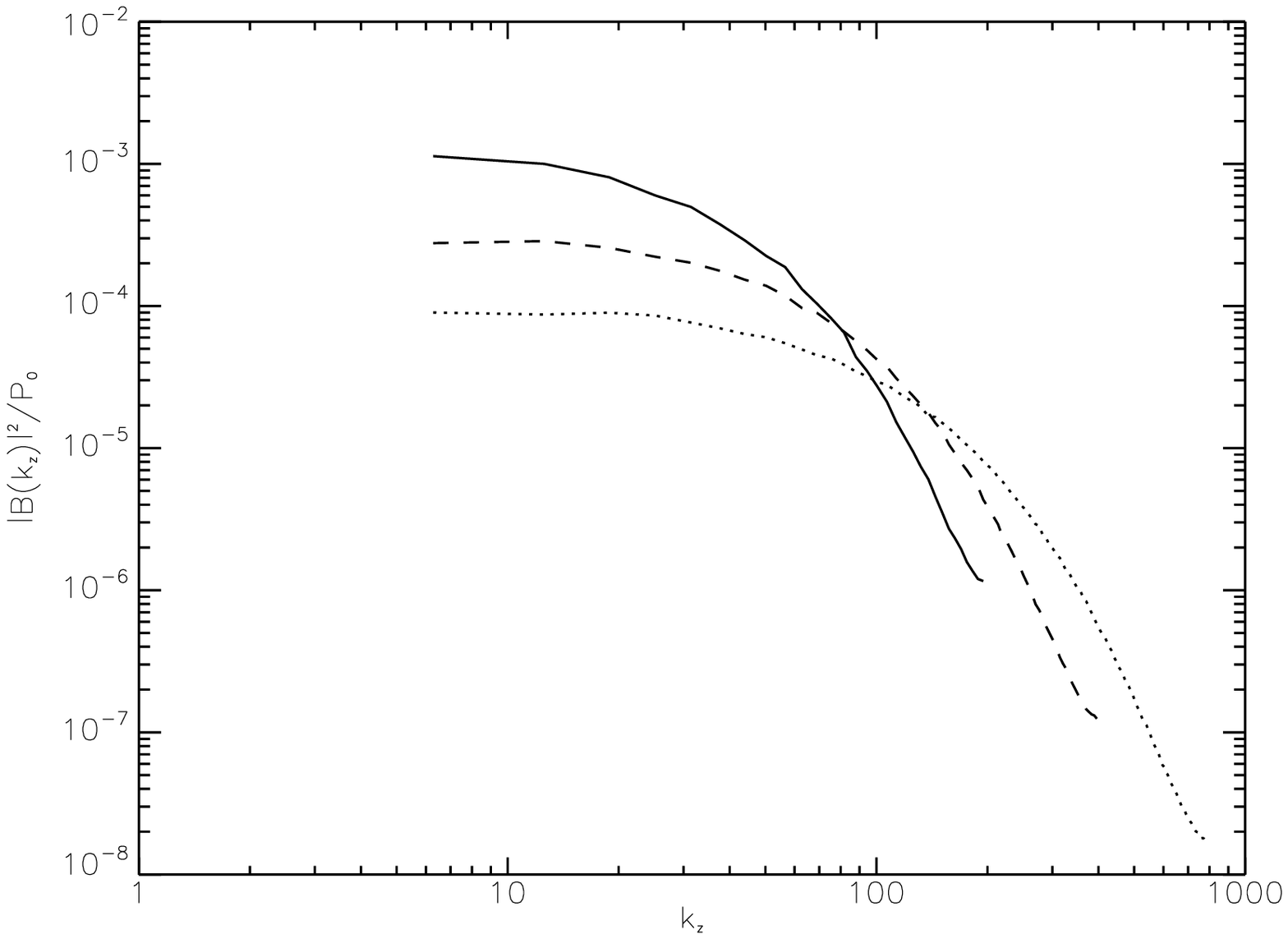}
\caption{Reduced power spectra of the kinetic energy ({\it upper panel}) and
  of the magnetic field ({\it lower panel}) in the z--direction. On
  both plots, the solid line corresponds to model STD64, the dashed
  line to model STD128 and the dotted line to model STD256. The 
  dot-dashed line on the upper panel shows the slope
  $k^{-11/3}$ expected in the standard Kolmogorov theory of incompressible
  hydrodynamic turbulence.In this and similar plots $k$ is expressed
  in units of $1/L$.}
\label{spectra_compar}
\end{center}
\end{figure}

Useful clues into the nature of MHD turbulence are usually provided by
power spectra. Here we compute the reduced power spectrum of kinetic
energy in the vertical direction, which we define as
\begin{equation}
E(k_z)=\frac{1}{2} \rho_0 |\bb{\tilde{v}}(k_z)|^2 \, ,
\label{kin spectrum}
\end{equation}
where
$|\bb{\tilde{v}}(k_z)|^2=|\tilde{v_x}(k_z)|^2+|\tilde{v_y}(k_z)|^2+|\tilde{v_z}(k_z)|^2$.
$\tilde{v_x}(k_z)$ is defined by 
\begin{equation}
{\tilde{v_x}}(k_z)=<\int_z v_x(x,y,z) e^{-i k_z z} dz> \, ,
\label{kin spectrum1}
\end{equation}
where $<.>$ stands for an average in the $x$ and $y$
directions. Similar definitions hold for $\tilde{v_y}(k_z)$ and
$\tilde{v_z}(k_z)$. In writing Eq.~(\ref{kin spectrum}), $\rho_0$
stands for the (conserved) mean density of the flow. Similar
expressions can be written to compute the reduced power
spectrum of magnetic energy.

Both spectra are represented in figure~\ref{spectra_compar} as a
function of $k_z$. The upper panel shows the kinetic energy
spectrum and the lower panel the magnetic energy spectrum. In both
panels, the results of model STD64 are shown using the solid line,
those of model STD128 are plotted with the dashed line and the dotted
line finally represents the results of model STD256. At all
resolutions, both kinetic and
magnetic energy spectra show features typical of turbulence: the
spectrum decreases with $k_z$, showing that there
is more energy at large scale. Note however the decreasing power
at large scales (both for the kinetic and magnetic energies) as
resolution is increased. This is because turbulent activity (or,
equivalently, angular momentum transport) decreases when resolution
increases and is in
agreement with the result of section~\ref{resol_effect_sec}. In the
upper panel, the dot--dashed line
  enables the results to be compared 
with the expected slope of a Kolmogorov
spectrum: $E(k_z) \propto k_z^{-11/3}$. There is no  identifiable region
 with the expected Kolmogorov slope, that  is maintained as the resolution
is increased, that can be seen  in 
 the computed spectra,  the best resolved calculation 
spanning  almost two orders of magnitude in $k_z$. For both the kinetic
and magnetic energies, the spectra consist of a flat part at large
scale, which grows in size as resolution is increased, and a
decreasing part probably governed by  numerical dissipation. In fact, these
spectra fail to show any sign of an inertial range building up as
resolution is increased. But is there any reason to expect these
simulations to show a clear inertial range?  Probably not. Because of the
MRI, the flow is a priori unstable at all realisable scales and
forcing and input is therefore
expected to occur all the way from the largest scale available in the
box down to the smallest MRI unstable scales (set by
numerical dissipation). At these scales a small scale
dynamo may also operate and even  transfer  some energy back 
to larger scales  \citep{Brummell98, Boldyrev05, 
 Ponty05}. Thus in our case there is no good reason
to suppose that any region of  Fourier space is
 expected
to be exclusively transferring kinetic or magnetic energy downward to
smaller scales,
as would be required for an inertial range to be observed. To
demonstrate this more clearly, we consider the properties of the Fourier
transformed induction equation in the next section.

In this context we comment that a situation where the MRI  leads to dynamo activity
differs from those such as occur when hydrodynamic phenomena such as the 
Rayleigh Taylor instability or Kelvin-Helmholtz instability produce turbulence.
In the case of the Rayleigh Taylor instability, the source of energy is confined
to the largest scales. Even though there are small scale  instabilities in
the linear regime, in the non linear regime these are overwhelmed by the
 advection process that results in the production of even smaller scales
where dissipation takes place \citep{Chertkov03}.
In the case of the Kelvin-Helmholtz instability, the situation is similar
with a source of instability occuring only at large scales when there is no
dynamo action  \citep{Nepveu85,Ryu00}.

\subsection{Equations}
\label{transfer_func}

We consider Eq.~(\ref{induct}) and decompose the
velocity as the sum of the mean shear flow and the turbulent velocity
field:
\begin{equation}
\bb{v}=\bb{V_{sh}}+\bb{v_{t}}
\end{equation}
where the mean shear flow is simply the $y$ and $z$ average of the $y$
component of the velocity.
\begin{equation}
\bb{V_{sh}}(x)=V_{sh} \bb{j} ={1\over L_y L_z} \int \int v_y(x,y,z)
 dy dz
\end{equation}
Using this decomposition, the right hand side of equation~(\ref{induct})
can be expanded and written as the sum of five terms:
\begin{equation}
\frac{\partial \bb{B}}{\partial t}  =  - V_{sh} \frac{\partial
  \bb{B}}{\partial y} + B_x \frac{\partial V_{sh}}{\partial x} \bb{j} -
  (\bb{v_t} \cdot \del) \bb{B} - (\del \cdot \bb{v_t})\bb{B} + (\bb{B}
  \cdot \del)\bb{v_t}
\label{induct_expand}
\end{equation}
where the dependence of first two terms on velocity is through
$V_{sh}$ only.  These describe advection by the mean flow and
stretching of the
radial magnetic field lines by the background shear. We  take the
Fourier transform of this equation and dot the resulting equation with
the complex conjugate of the Fourier transform of 
${\bb{B}}.$ 
Denoting the later by 
$\tilde{\bb{B^*}}(\bb{k})$ and noting that
\begin{equation} {\tilde{\bb{B}}}(\bb{k})= \int \bb{B}(\bb{x})\exp{(-i \bb{k}\cdot\bb{x}})d^3{\bb{x}},\label{FT} \end{equation}
 defines a finite  Fourier transform,
we obtain an 
equation governing the evolution of the magnetic energy density in Fourier
space in the form
\begin{equation}
\frac{1}{2}\frac{\partial |\tilde{\bb{B}}(\bb{k})|^2}{\partial t} = A
+ S + T_{bb} + T_{divv} + T_{bv} \, ,
\label{fourier_induct}
\end{equation}
where 
\begin{eqnarray}
A &=& -{\cal R}e\left[\tilde{\bb{B^*}}(\bb{k}) \cdot \int \int \int  V_{sh} \frac{\partial
  \bb{B}}{\partial y}  e^{-i\bb{k}.\bb{x}} d^3\bb{x}\right] \\
S &=& +{\cal R}e\left[\tilde{B_y^*}(\bb{k}) \cdot \int \int \int B_x
  \frac{\partial V_{sh}}{\partial x}  e^{-i\bb{k}.\bb{x}} d^3\bb{x}\right] \\
T_{bb} &=& -{\cal R}e\left[\tilde{\bb{B^*}}(\bb{k}) \cdot \int \int \int  
  [(\bb{v_t} \cdot \del) \bb{B} ] e^{-i\bb{k}.\bb{x}} d^3\bb{x}\right] \\
T_{divv} &=& -{\cal R}e\left[\tilde{\bb{B^*}}(\bb{k}) \cdot \int \int \int  
  (\del \cdot \bb{v_t})\bb{B}  e^{-i\bb{k}.\bb{x}} d^3\bb{x} \right]\\
T_{bv} &=& +{\cal R}e\left[\tilde{\bb{B^*}}(\bb{k}) \cdot \int \int \int  
  [(\bb{B} \cdot \del)\bb{v_t}]  e^{-i\bb{k}.\bb{x}} d^3\bb{x}\right] \, 
\end{eqnarray}
where ${\cal R}e$ denotes the real part is to be taken. Note that we
applied the remap procedure described in \citet{hawleyetal95} to
account for the shear when computing the Fourier transform in the $x$
direction.

During the saturated phases of the previously described simulations, the
time derivative of $|\tilde{\bb{B}}(\bb{k})|^2$  should vanish on average.
In simulations of the MRI \citep{hawleyetal95,brandenburgetal95}
 it is quite generally found that quantities are stretched out
in the direction of the shear and thus the length scales in the $x$
and $z$ direction associated with the saturated state 
are significantly smaller than the length scale in the $y$ direction.
See for example Figure 4 of \citet{hawleyetal95}.
 Therefore we shall for simplicity
consider only the particular Fourier plane defined by $k_y=0$ and then
 we get
$A=0$. Under these  assumptions, Eq.~(\ref{fourier_induct})
simplifies to
\begin{equation}
S + T_{bb} + T_{divv} + T_{bv}=0 \, .
\label{master_tr}
\end{equation}
In this equation, $S$ describes how the background shear creates the
$y$ component of the magnetic field, $T_{bb}$ is a term that accounts
for magnetic energy transfer toward smaller scales, $T_{divv}$ is due to
compressibility and $T_{bv}$ describes how magnetic field is created
due to field line stretching by the turbulent flow. Each of these
terms now depends on the wavenumber $\bb{k}=(k_x,k_z)$. To improve the
statistics, we average them on shell of given modulus $k=|\bb{k}|$ as well as
over time. 

Of course, in a real simulation of the type considered here, magnetic energy is damped
by numerical dissipation. Therefore, a more realistic equation than
Eq.~(\ref{master_tr}) would be
\begin{equation}
S + T_{bb} + T_{divv} + T_{bv} + D_{num}=0 \, .
\label{master}
\end{equation}
where $D_{num}$ accounts for the numerical dissipation (of course 
'realistic' physical dissipation can be treated in a similar way, see
paper II). In
the following section, we study the balance between the various terms of
Eq.~(\ref{master}) for models STD64, STD128 and STD256.

\subsection{Results}

\begin{figure}
\begin{center}
\includegraphics[scale=0.45]{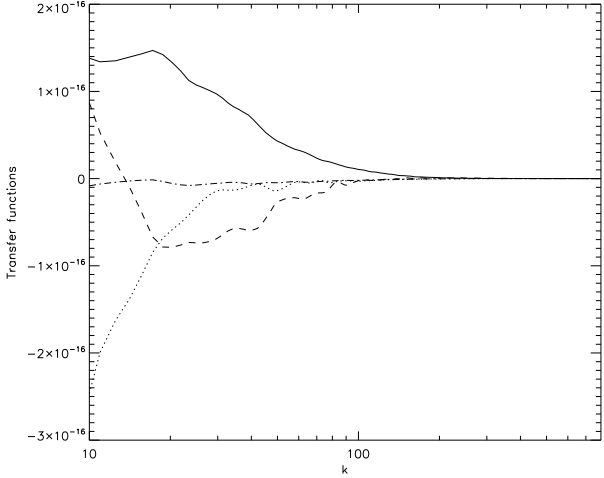}
\caption{Plots of the functions $S$ ({\it solid line}), $T_{bb}$
  ({\it dashed line}), $T_{divv}$ ({\it dotted line}) and $T_{bv}$
  ({\it dotted--dashed line}) as functions of $k$ for model STD256.
   These are averaged in time and Fourier space as described in the text.
   Note that $S$ is
  positive at all scales, which is simply describing the production of
  $B_y$ by the background shear.}
\label{tot_transfer_256}
\end{center}
\end{figure}

\begin{figure}
\begin{center}
\includegraphics[scale=0.45]{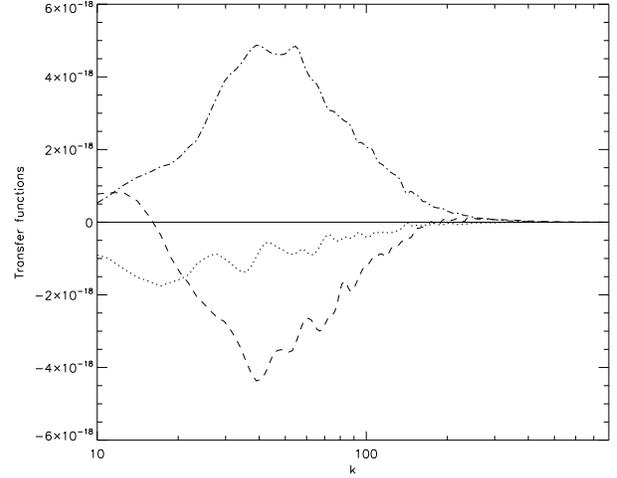}
\caption{Same as figure~\ref{tot_transfer_256}, but for the poloidal
  part of the magnetic energy. The term $T_{bv}^P$ is positive at all
  scales, indicating
  that MHD turbulence is forced by the MRI from the largest scale
  available in the simulation box down to the dissipative scale.}
\label{polo_transfer_256}
\end{center}
\end{figure}

\begin{figure}
\begin{center}
\includegraphics[scale=0.45]{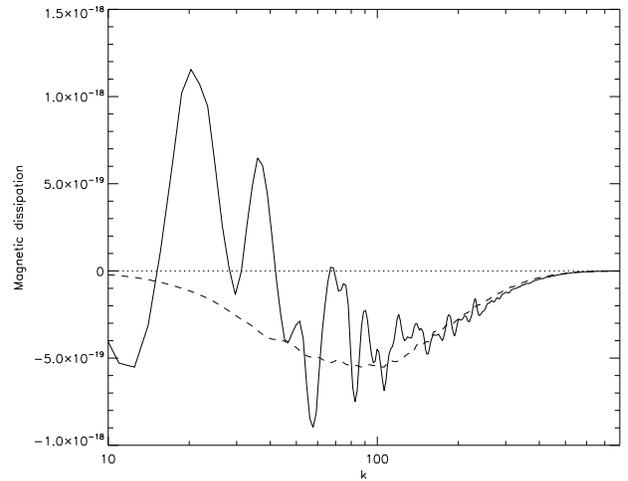}
\caption{The solid line shows the variation of the numerical dissipation
  $D_{num}^P=-(T_{bb}^P+T_{divv}^P+T_{bv}^P)$ with wavenumber $k$ for model
  STD256. It is compared, using the dashed line, with the spectrum $D_{res}$
  that would result from a purely resistive dissipation corresponding to
  a magnetic Reynolds number $Re_M=10^5$ (see the text for the
  definition of $Re_M$). The thin solid line simply indicate the
  location of the zero on the plot. Note the good agreement between $D_{num}^P$
  and $D_{res}$ at small scales, while they disagree significantly at large
  scales, showing that ZEUS numerical dissipation departs from a pure
  physical resistivity.}
\label{dissip_transfer_256}
\end{center}
\end{figure}

As model STD256 is the most detailed of our runs in term of
resolution, we start by describing the results we obtained in this
case before comparing with the other simulations. As mentioned above,
the results of model STD256 were averaged in time to improve the
statistics. We used $20$ dumps spanning about $60$ orbits from $t=45$
until the end of the simulation. Figure~\ref{tot_transfer_256} plots the
four terms appearing in Eq.~(\ref{master_tr}) versus $k$.
 The solid,
dashed, dotted and
dotted--dashed lines respectively correspond to the terms $S$,
$T_{bb}$, $T_{divv}$ and $T_{bv}$. Only the first is
positive, while the other three terms are mainly negative, except for
$T_{bb}$ which is positive at the largest scale of the box. The term
$T_{divv}$, accounting for compressibility, is also seen to reach
significant values, probably because of the presence a strongly
nonlinear waves in these simulations
\citep{gardiner&stone05b,papaloizouetal04}. The large and positive
value of the $S$ term simply shows that
$B_y$ is created at all scales by the background shear. For the MRI to
be a proper dynamo, however, there has to be a way through which poloidal
magnetic energy is regenerated from this toroidal field. To study that
mechanism, we redo the
analysis presented above but concentrate on the
poloidal part of the magnetic field $\bb{B_p}=(B_x,0,B_z)$ rather than
on the full magnetic field $\bb{B}$. In that case, both $S$ and $A$
vanish and under the assumption that MHD turbulence is in steady
state, Eq.~({\ref{master}) reduce to 
\begin{equation}
T_{bb}^P + T_{divv}^P + T_{bv}^P + D_{num}^P=0 \, ,
\label{master_polo}
\end{equation}
where we have now
\begin{equation}
T_{bb}^P = {\cal R}e\left[\tilde{\bb{B_p^*}}(\bb{k}) \cdot \int_x \int_y \int_z  
  (\bb{v_t} \cdot \del) \bb{B_p}  e^{-i\bb{k}.\bb{x}} d^3\bb{x}\right]
\end{equation}
with corresponding expressions for the other terms in
Eq~({\ref{master_polo}}). The results
we obtained for model STD256 are plotted on 
figure~\ref{polo_transfer_256}, with the same conventions 
 as figure~\ref{tot_transfer_256}. They indicate
 that
$T_{bv}^P$ is positive for all $k$ through   turbulent velocity fluctuations
 creating poloidal magnetic field through field line
stretching. $T_{bv}^P$ reaches its maximum at $k \sim 40$,
which corresponds to roughly $1/8$th the size of the box (or $32$ grid
cells at this resolution). Nevertheless there are non negligible
contributions from the largest scale available in the box all the way
down to $k \sim 200$, which corresponds to only a few grid cells. This
is an indication that the MRI is forcing the flow at all available
scales and explains why the power spectra
presented in section~\ref{power_spec_sec} fail to show any inertial
range. 

Another application of the above analysis is to provide information
about the dissipative properties of ZEUS. Indeed, because of
Eq.~(\ref{master}), the sum of the three terms represented on
figure~\ref{polo_transfer_256} has to be balanced by the
numerical dissipation. If that dissipation was exactly equivalent to a
resistive process having a resistivity $\eta$, its spectrum
$D_{res}(k)$ would be given by
\begin{equation}
D_{res}(k)=\eta k^2 |\tilde{\bb{B_p}}(k)|^2 \, .
\end{equation}
In Figure~\ref{dissip_transfer_256} we plot $D_{num}^P$ ({\it solid
  line}) and $-D_{res}$ ({\it dashed line}). In the later case we
  chose $\eta$ in order to get a good fit between both curves. At large
  $k$ (or small scales), it is seen that $-D_{res}$ is in good
  agreement with $D_{num}^P$. This  fit can be used to
  estimate the
  numerical resistivity $\eta$ of the code (at this particular
  resolution for small scales and for this specific flow). This
  translates into an
  effective magnetic Reynolds number $Re_M=c_0H/\eta$ which is of the order
  of $10^5.$ This equivalence has to be used with caution, however, as both 
  $D_{num}^P$ and $-D_{res}$ deviate significantly  for 
  values of $k$ smaller than $80$. A contribution to this difference may
  arise from
  the poorer statistics available at the largest scale of the
  simulations (and particularly to non negligible contribution of the
  time derivative term in Eq.~(\ref{fourier_induct}) at these scales),
  but it is also likely to be  due to the fact that numerical
  dissipation  cannot be simply  described  as arising from a diffusion process,
  at least at large scales.  Also, it should be
  noted that the maximum amplitude of $D_{num}^P$ (and therefore the
  location where most of the dissipation takes place) occurs at $k \sim
  70$--$100$, at which point the term $T_{bv}^P$ is still very
  significant (see figure~\ref{polo_transfer_256}). This is why the
  saturated state of MRI driven MHD turbulence depends on
  resolution. Based on this analysis, we would therefore predict that
  increasing the resolution by another factor of two would  give a
  different saturated rate of angular momentum transport. 

\begin{figure*}
\begin{center}
\includegraphics[scale=0.4]{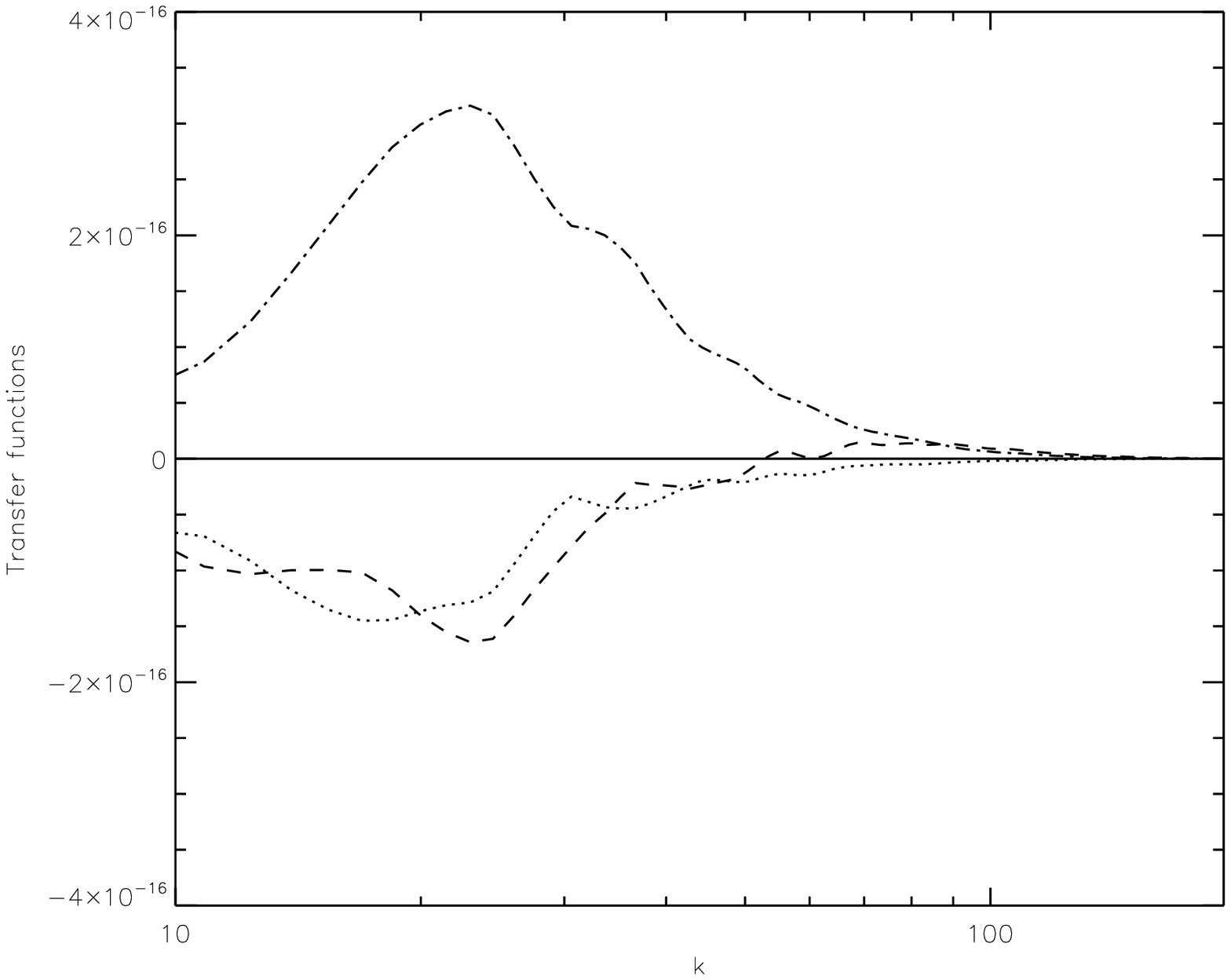}
\includegraphics[scale=0.4]{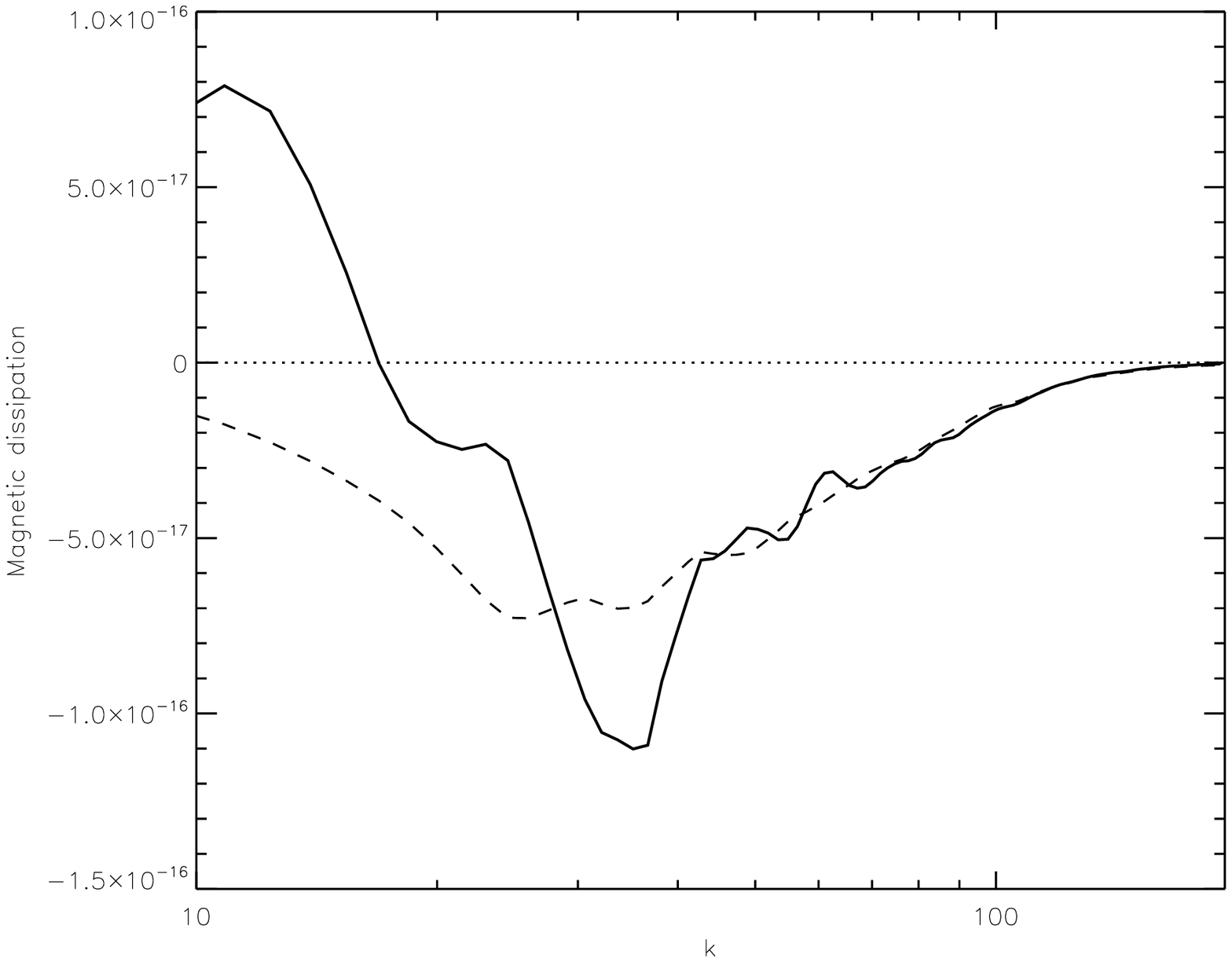}
\caption{The left panel is the same as figure~\ref{polo_transfer_256},
  but computed using results of model STD64. Likewise, the right panel
  is the same as figure~\ref{dissip_transfer_256} applied to model
  STD64. The dashed line use a resistivity $\eta$ such that $Re_M=10^4$.}
\label{transfer_64}
\end{center}
\end{figure*}

\begin{figure*}
\begin{center}
\includegraphics[scale=0.4]{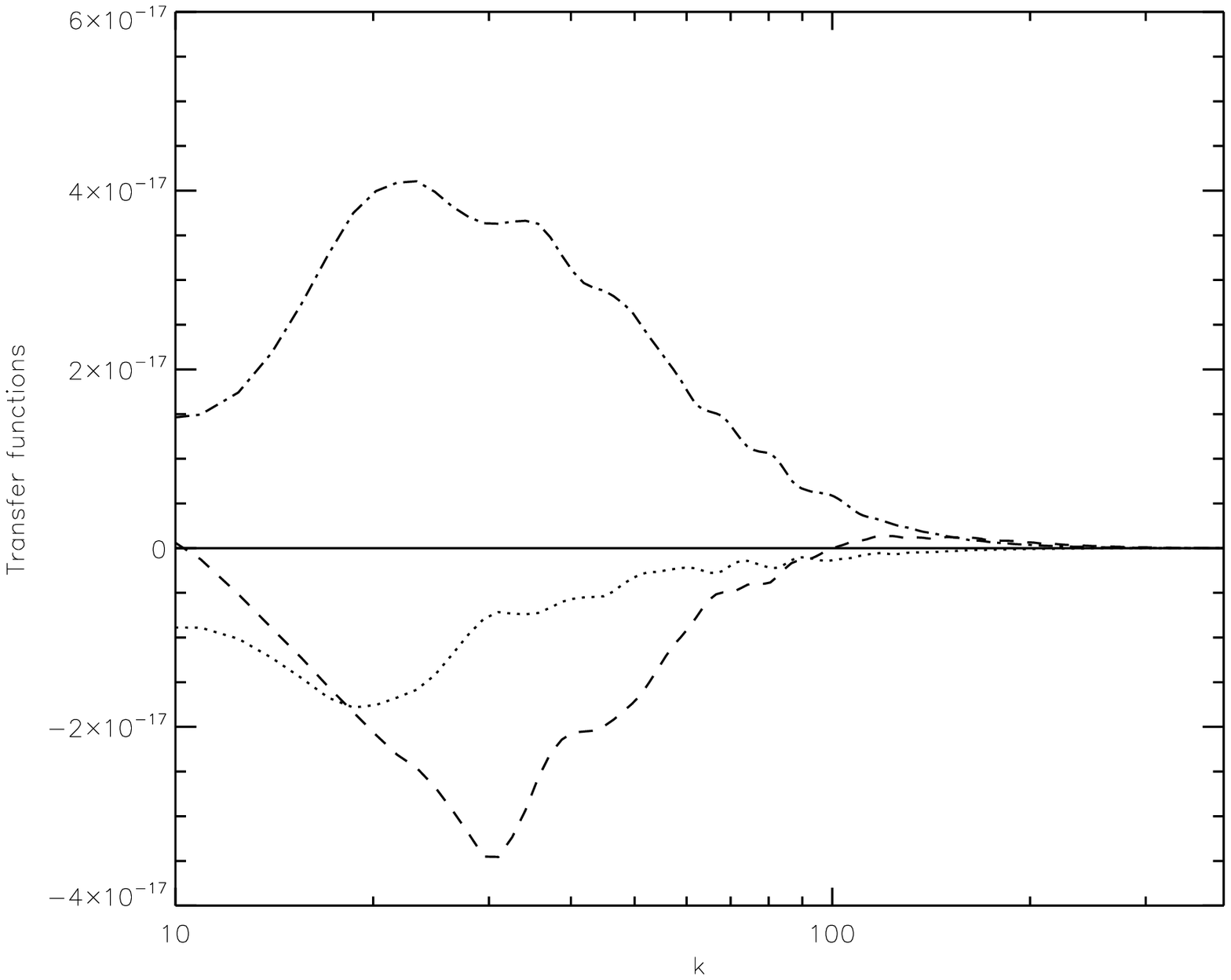}
\includegraphics[scale=0.4]{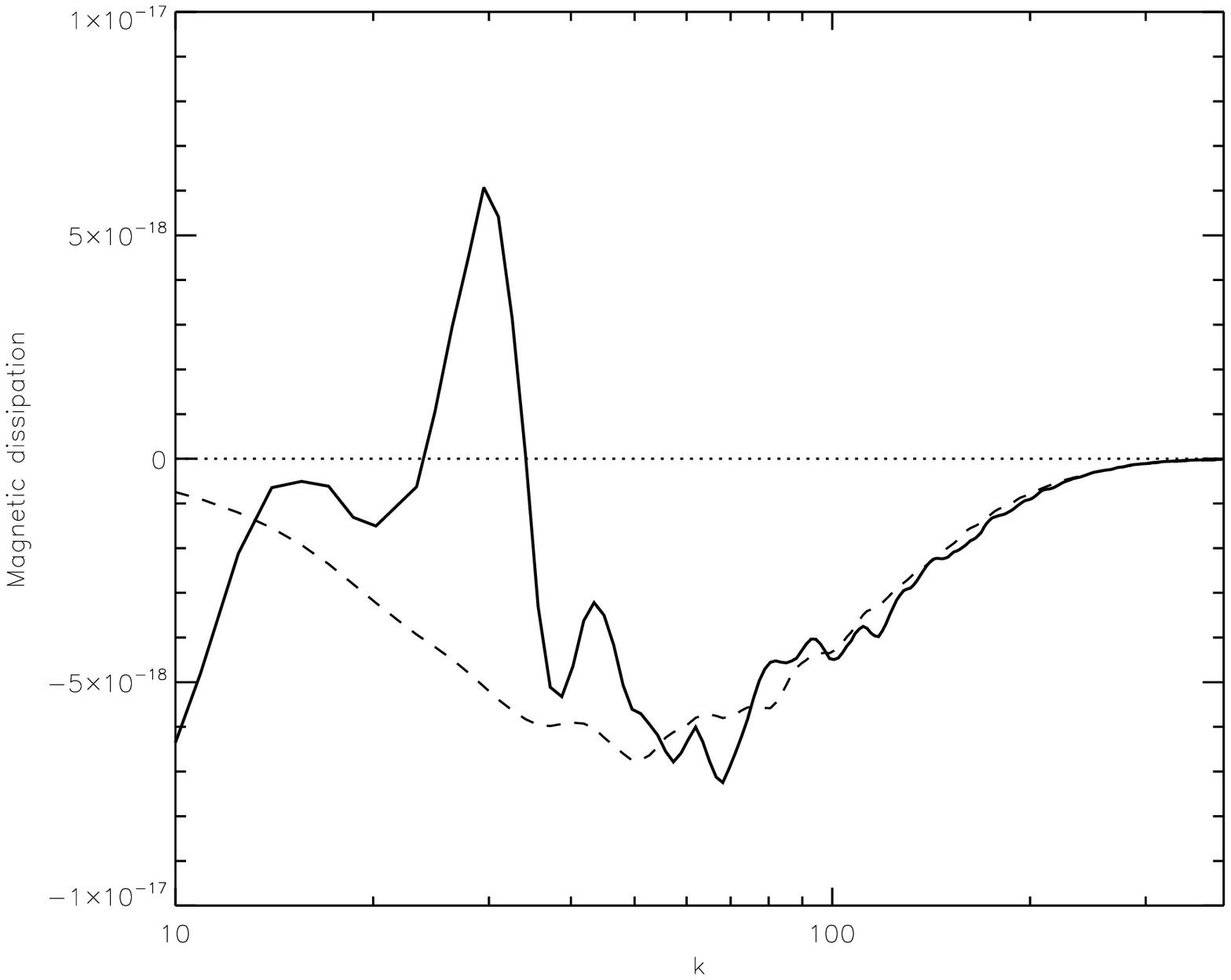}
\caption{Same as figure~\ref{transfer_64}, but for model STD128. On
  the right panel, the value of $\eta$ used to fit the numerical
  results is such that $Re_M=3 \times 10^4$.}
\label{transfer_128}
\end{center}
\end{figure*}

In order to further investigate the effect of resolution, we repeated
the previous analysis for models STD64 and STD128. For model STD64, we
used about $90$ dumps regularly spaced in time during the $1000$
orbits of the simulations to average the results. For model STD128, we
used $60$ dumps that cover the last $200$ orbits of the
simulations. The results are summarised in figures~\ref{transfer_64}
and \ref{transfer_128} respectively. On both figures, the left panel
is the equivalent of figure~\ref{polo_transfer_256}. These confirm the
results obtained using model STD256. The MRI forces the flow at all
available scales in the computational box. The right panels of
figures~\ref{transfer_64} and \ref{transfer_128} are equivalent to
figure~\ref{dissip_transfer_256} but  for models STD64 and STD128
respectively. Both confirm that small scale dissipation in ZEUS  is
similar to that provided by a  physical resistivity, with magnetic
Reynolds numbers of the order of $10^4$ and $3 \times 10^4$
respectively. However, as also seen for model STD256, numerical
resistivity departs from a physical resistivity at large scale in a
way that depends on resolution (note that
the amplitude of the oscillations seen in plotting the numerical
dissipation is reduced compared to model STD256,
which is illustrating the fact that the statistics are improved when
the simulation is integrated longer; as a result, the deviation of the
numerical dissipation from a purely Laplacian process appears more
solid). Both model STD64 and STD128 indicate that the numerical scheme in
ZEUS is such that the numerical resisitivity could be negative at
large scale. This is also suggested by model STD256 (see
figure~\ref{dissip_transfer_256}) although poor statistics make that
conclusion less clear in that case. This possible antidiffusive
behavior of ZEUS was in fact already pointed out by \citet{falle02} for
1D shock calculations and has recently been observed
to occur when studying the propagation of torsional Alven waves
 \citep{lesaffre&balbus07}. It is not clear whether this is intrisic 
to ZEUS numerical scheme, to peculiarities introduced by the shearing
box boundary conditions, or to a combination of both.

\section{Discussion}
\label{discussion_section}

\subsection{The magnetic Prandtl number for ZEUS}

\begin{figure}
\begin{center}
\includegraphics[scale=0.45]{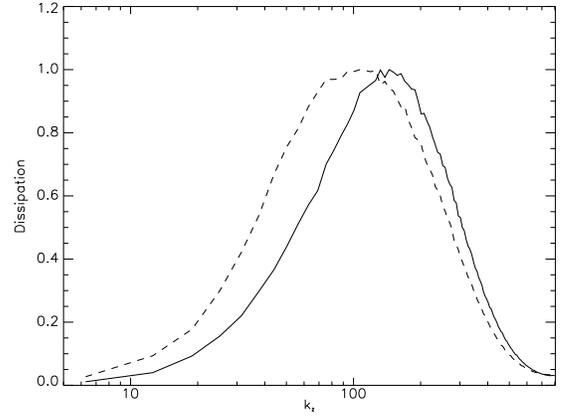}
\caption{Reduced power spectrum of the velocity and magnetic field
 averaged over time in model STD256. 
  The plots represent the quantities
    $k_z^2|\bb{{\tilde B}}(k_z)|^2 $ 
  ({\it solid line}) 
     and
     $k_z^2|\bb{{\tilde v}}(k_z)|^2$
    ({\it dashed line})      
    which would measure the rate of
  dissipation for constant diffusivities. 
  The dashed curve peaks at smaller $k_z$ than the solid
  curve indicating that the hydrodynamic dissipation length is larger
  than its MHD counterpart. 
  This in turn indicates that the numerical Prandtl
  number of ZEUS is larger than $1$.
\label{pm_num_256}}
\end{center}
\end{figure}

\begin{figure}
\begin{center}
\includegraphics[scale=0.25]{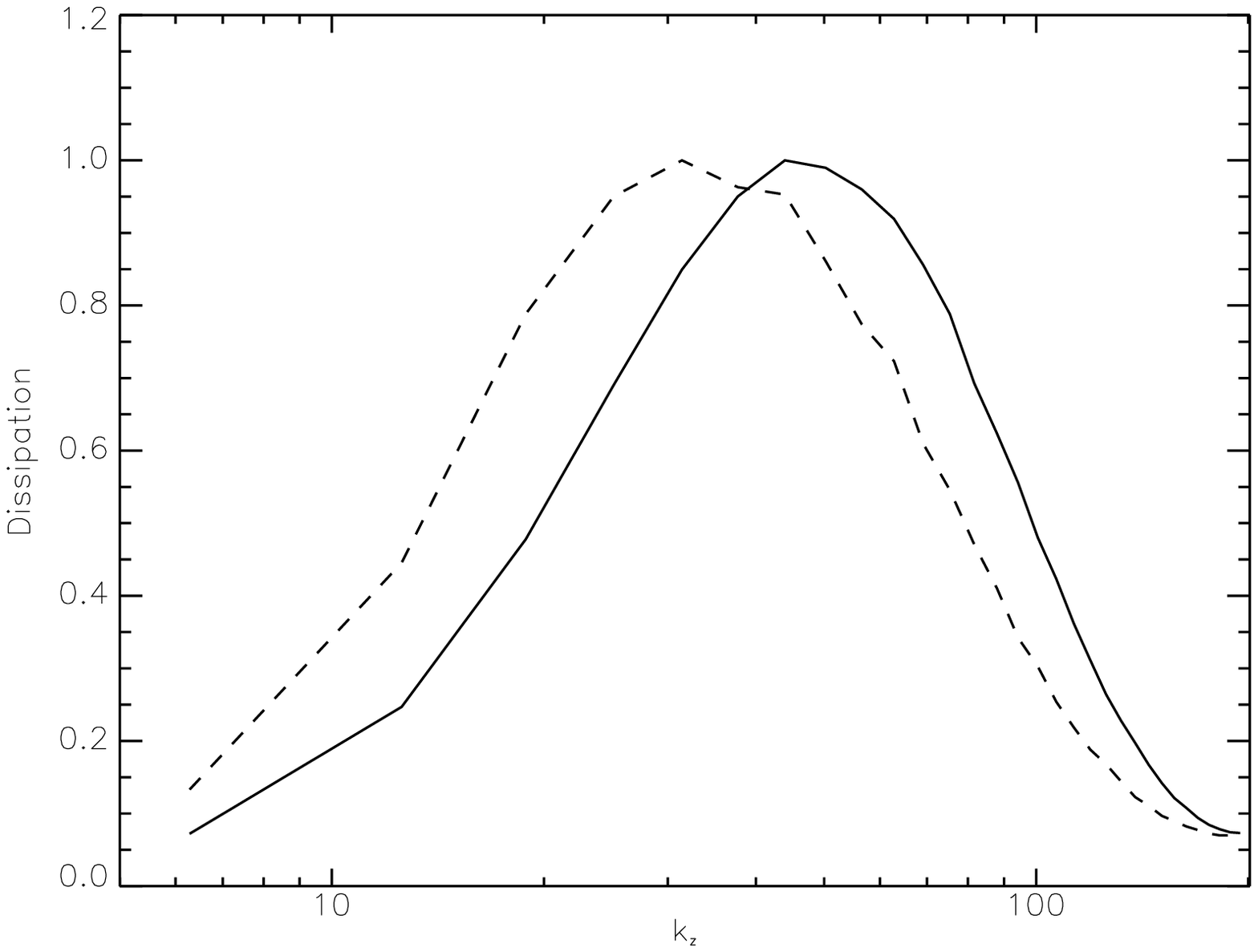}
\includegraphics[scale=0.25]{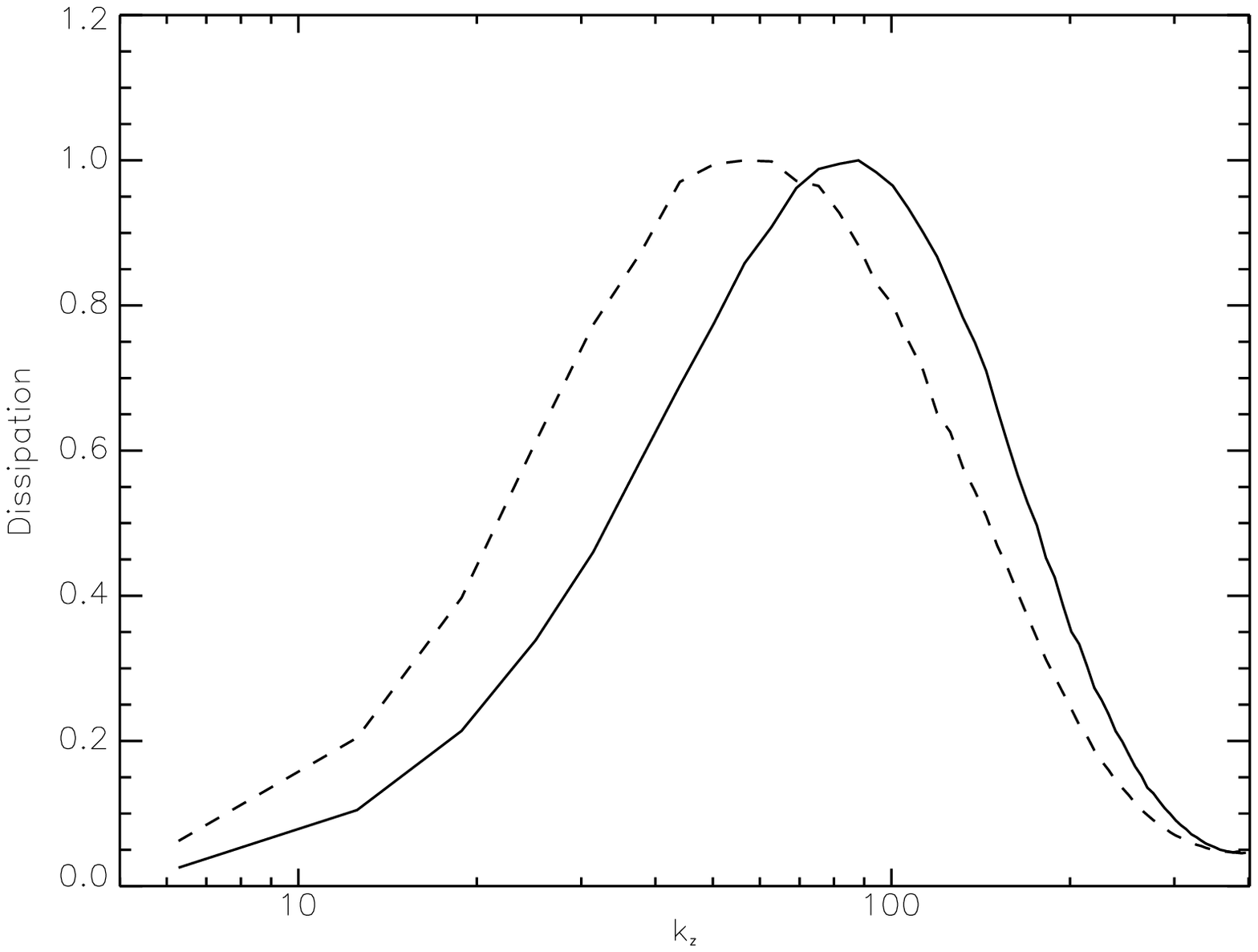}
\caption{Same as figure~\ref{pm_num_256}, but for model STD64 ({\it
    left panel}) and STD128 ({\it right panel}). Both confirm that the
    numerical Prandtl number for ZEUS is larger than unity, in
    agreement with the results obtained with model STD256.}
\label{pm_num_res}
\end{center}
\end{figure}

The previous analysis explains why numerical simulations performed with
ZEUS fail to converge when the resolution is increased. It also leads to an
estimate of the numerical resistivity associated with such a
calculation. However,  the numerical effective kinematic viscosity $\nu,$
associated with the momentum equation, has not been discussed.
On general grounds one expects this to be of the
same order of magnitude as the numerical resistivity. To be more
quantitative, it would be necessary to perform an analysis similar to that presented
in section~\ref{transfer_func}, but for the evolution of the kinetic energy
per unit mass, in order to estimate an effective  $\nu$, which
could in turn be used to obtain a measure of the numerical magnetic Prandtl number
$Pm$ for ZEUS:
\begin{equation}
Pm=\frac{\nu}{\eta} \, .
\end{equation}
Such a procedure is however  complex and beyond the scope of this
paper.

As an alternative, we obtain an indication of  the value of $Pm$ by
 considering  ${\cal D}_{res}=k_z^2 |\tilde{\bb{B}}(k_z)|^2$ and
 ${\cal D}_{vis}=k_z^2|\tilde{\bb{v}}(k_z)|^2$.  Here
 $\tilde{\bb{B}}(k_z)$ is, to within an ignorable constant factor, the
 quantity associated with the magnetic field corresponding to
 $\tilde{\bb{v}}(k_z)$ for the velocity  field defined through
 equations (\ref{kin spectrum}-\ref{kin spectrum1}).  It may also be
 found by evaluating the Fourier transform (\ref{FT}) for $k_x=k_y =0.$

These quantities would be proportional to the numerical  resistive and
hydrodynamic  dissipation for $k_x=k_y=0,$ if the later could be
described by a simple diffusion process. As we demonstrated in the
previous section, this is not generally the case, but it appears to be
reasonable for the smallest scales in the case of resistive
dissipation. In order to obtain an indication of the effective value
of $Pm,$ we shall make the very reasonable  assumption that the
numerical viscous dissipation at small scales can be described in the
same way.  This is expected because there are no strong shocks and the
order of the finite difference scheme is the same for the induction
equation and the equation of motion.  Also  there are no added
hyperdiffusive terms, which would require a dependence on higher
powers of $k_z$ in ${\cal D}_{res}$ and ${\cal D}_{vis}$.

Both ${\cal D}_{res}$ and ${\cal D}_{vis}$ are plotted in
figure~\ref{pm_num_256} for model STD256 using solid and dashed lines
respectively. Both curves are time averaged between $t=50$ orbits
until the end of the simulation and are normalised by their maximum
values. Figure~\ref{pm_num_256} shows that ${\cal D}_{vis}$ peaks at a
wavenumber $k_z^{\nu}$ which is smaller than the wavenumber
$k_z^{\eta}$ at which ${\cal D}_{res}$ reaches its maximum. These
peaks should indicate the scale at which most of the dissipation
occurs. In other words, they can be used for order of magnitude
estimates of both the viscous and the resistive lengths $l_{\nu}$ and
$l_{\eta}$. From figure~\ref{pm_num_256}, we find $k_z^{\nu} \sim 100$
and $k_z^{\eta} \sim 150$. This means that $Pm$ is of order unity, and
probably biased toward values larger than one. Indeed, since
$k_z^{\nu}$ is smaller than $k_z^{\eta}$, it follows that the viscous
length is larger than the resistive length, or that numerical
viscosity should be larger than numerical resistivity (see also the
discussion in paper II).

Of course, there is significant uncertainty associated with the above
estimate and with the method we used to derive it, but
the fact that $Pm$ is larger than $1$ appears to be  solid. It is
further confirmed by the results shown in figure~\ref{pm_num_res},
which is the same as  figure~\ref{pm_num_256} but for model STD64
({\it left panel}) and STD128 ({\it right panel}). In both cases, the
dashed curve peaks at smaller wavenumbers than the solid curve,
indicating that the viscous length is larger than the resistive
length, in agreement with the discussion above.

\subsection{Scaling arguments}
\label{scaling_section}

\begin{figure}
\begin{center}
\includegraphics[scale=0.5]{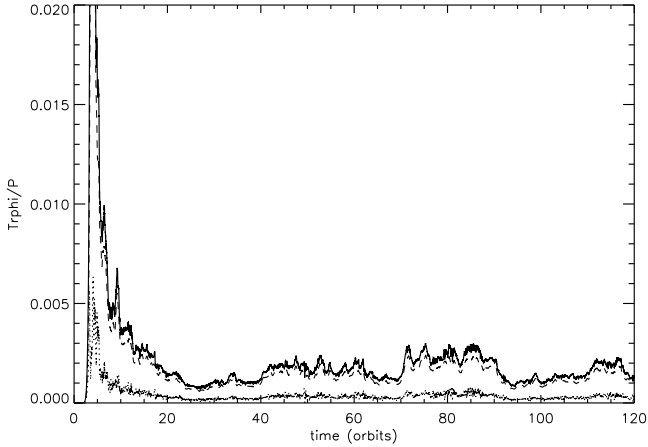}
\caption{Time history of the stresses for the run STD64a. 
The dotted
  curve corresponds to the Reynolds stress, the dashed curve
  corresponds to the Maxwell stress and  the solid curve  gives the
  sum of the two. All of these are normalised by the initial thermal pressure. }
\label{history}
\end{center}
\end{figure}

\begin{figure}
\begin{center}
\includegraphics[scale=0.5]{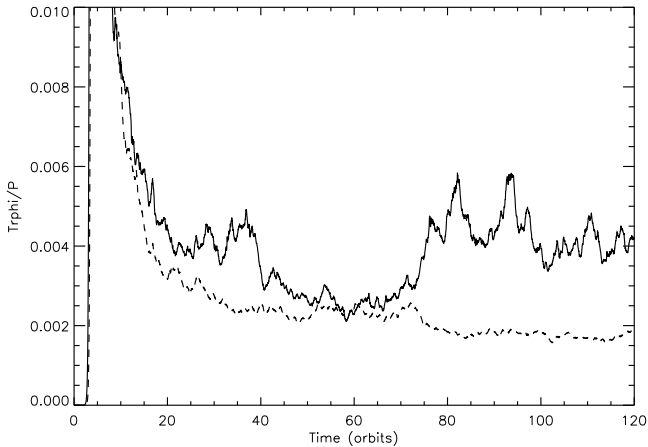}
\caption{Time history of the Maxwell stress for the runs LB64 ({\it solid
    line}) and LB128 ({\it dashed line}). For the former, time average
    between $t=40$ and the end of the run gives $\alpha=5.2
    \times 10^{-3}$ while $\alpha=3.2 \times 10^{-3}$ for the
    latter. Thus angular momentum transports decreases when resolution is increased.}
\label{hist_bigbox}
\end{center}
\end{figure}

We note that the shearing box equations and boundary conditions
may be transformed to a dimensionless
representation. This is the case either, when the evolution
is governed by partial differential equations, or by the finite difference
equations of a numerical scheme.
The transformation is performed by choosing $L$, the box size,
$\Omega^{-1}$ and $\rho_0 L^3$ 
as the units
of length,  time and mass  respectively.
The resulting equations then depend only on the dimensionless quantities
$H/L,$ $h/L$ and $C,$ where $h$ is the grid spacing and $C$ denotes the
Courant number (assuming a fixed aspect ratio for the grid cells).
The unit of magnetic field is then $ \sqrt{L^2\Omega^2\rho_0}.$
Consequently we expect any one of the stress parameters to have the scaling
\begin{equation}
 \alpha  \propto  \left(L/H \right)^2F(h/L,L/H,C),
\end{equation}
where $F$ is  some unspecified function.
For our calculations $C=1/2$ is fixed while runs STD64, STD128, and STD256
which have fixed $L$ and $H$ indicate that F is $\propto h$ under those
constraints and within the range of $h$ considered. Then we may write
\begin{equation}
 \alpha  \propto  (Lh/H ^2) G(L/H).
\label{alpha_scaling}
\end{equation}
For some unspecified function $G.$ We have investigated  the form of
$G$ by performing simulation STD64a. This has L reduced by a factor of two 
and h increased by a factor of two when compared to STD256. Thus if $G(L/H)$
were constant, the stress parameters should be the same for the two runs.
In fact the data show that on average the stresses were about 80
percent  larger in STD64a (see figure~\ref{history} and
table~\ref{zero_diss_prop}, respectively giving the stresses time
histories and averaged values). On the other hand the stresses showed
significantly 
stronger time variability in that case but with a base level comparable to that
in STD256. These results indicate that qualitative as well as quantitative
changes occur when dimensionless parameters are varied.
These differences may be due to, for example, a variation of the
importance of compressibility as has been considered by
\citet{sanoetal04}, or a variation of the effective Prandtl
number. The importance of the Prandtl number as determined by the
physical diffusion coefficients when these determine the form of the
saturated state is considered in a companion paper.

It is also of interest to ask whether the scaling of $\alpha$
  with resolution described in the present paper and expressed by
  Eq.~(\ref{alpha_scaling}) still holds for larger boxes. This could
  have important implications for global simulations. Thus we
  performed two additional simulations with a box of size
  $(L_x,L_y,L_z)=(2H,2\pi H,H)$. The first one, labelled LB64, has a
  resolution $(N_x,N_y,N_z)=(128,200,64)$, which amounts to $64$ grid
  points per scale height. In the second, LB128, the resolution is
  doubled. Figure~\ref{hist_bigbox} illustrates the results through the
  time history of the Maxwell stress (the solid line corresponds to
  model LB64 and the dashed line to model LB128). Again, we found a
  significant decrease of the turbulent activity as resolution is
  increased: $\alpha=5.2 \times 10^{-3}$ for model LB64 and $2.8
  \times 10^{-3}$ for model LB128. This tends to indicate that the
  results we present in this paper could extend to global disk
  simulations. Unfortunately the very high computational demands
  associated with these simulations precludes extensive studies at
  this time.

\section{Conclusion}
\label{conclusion_section}

In this paper, we have shown that angular momentum transport induced
by MHD turbulence decreases when the resolution is increased in numerical
simulations performed with ZEUS in a shearing box in the
  absence of net magnetic flux. We have shown that this is due to the
MRI forcing the flow at all scales, including those at which
dissipation takes
place. There is enough energy  at these smallest scales 
to affect mean stresses in the saturated state.
 All our results, taken together, demonstrate that it is
important to use explicit diffusion coefficients that are
large enough to produce more dissipation than numerical effects
in local numerical simulations of MHD turbulence with zero net
  flux performed with a finite difference code like ZEUS at currently
feasible resolutions. Recent numerical simulations of MHD
  turbulence in the shearing box in the presence of an imposed
  magnetic flux showed that $\alpha$ also depends on physical
  dissipation in that case \citep{lesur&longaretti07}, as it was found
  that it increases as the ratio of kinematic viscosity to magnetic
  diffusivity does.

We note that there is no reason why this state of affairs
should not apply  
when using  Godunov codes like ATHENA \citep{gardiner&stone05a} or
RAMSES \citep{teyssier02,fromangetal06} or other codes making use of
hyperviscosity to stabilise the numerical scheme, like the PENCIL code
\citep{brandenburg&dobler02} for example. 

A study of the effects of magnetic diffusivity and kinematic viscosity
on local numerical simulations of MHD turbulence with zero net flux is
the subject of a companion paper.

Finally we comment that the results of this paper apply to the very simple
computational set up of a local unstratified shearing box with zero net flux
and for a restricted domain in parameter space.
 They should not be applied to more complex
stratified or global simulations which will require separate studies.
Neither should they be extrapolated beyond the parameter ranges considered.

\section*{ACKNOWLEDGMENTS}

We thank Geoffroy Lesur, Gordon Ogilvie, Fran{\c c}ois Rincon and Alex
Schekochihin for useful discussions. The simulations presented in this
paper were performed using the Cambridge High Performance Computer
Cluster Darwin and the UK Astrophysical Fluids Facility (UKAFF). We
thank the referee, Jim Stone, for helpful suggestions that
significantly improved the paper.

\bibliographystyle{aa}
\bibliography{7942}

\end{document}